\documentclass[aps,preprint,nofootinbib]{revtex4}
\usepackage{amsfonts}
\usepackage{amsmath}
\usepackage{amssymb}
\usepackage{relsize}
\usepackage{graphicx}
\usepackage{pdfpages}
\usepackage{subfigure}
\usepackage{float}
\usepackage[caption = false]{subfig}
\usepackage{color}
\usepackage{multirow}

\def\correspondingauthor{\footnote{Corresponding author. E-mail address: liverts@phys.huji.ac.il}}
\begin{document}

\title{Electrons on sphere in the helium-like atomic systems}

\author{Evgeny Z. Liverts\correspondingauthor{}}
\affiliation{Racah Institute of Physics, The Hebrew University, Jerusalem 91904, Israel}



\begin{abstract}
The properties of a special configuration of a helium-like atomic system, when both electrons are on the surface of a sphere of radius $ r $, and angle $\theta$ characterizes their positions on sphere, are investigated.
Unlike the previous studies, $r$ is considered as a quantum mechanical variable but not a parameter.
It is important that the "electrons-on-sphere" and the "collinear" configuration are coincident in two points.
For $\theta=0$ one obtains the state of the electron-electron coalescence, whereas the angle $\theta=\pi$ characterizes
the $\textbf{e-n-e}$ configuration when the electrons are located at the ends of the diameter of sphere with the nucleus at its center.
The Pekeris-like method representing a fully three-body variational technique is used for the expedient calculations.
Some interesting features of the expectation values representing the basic characteristic of the  "electrons-on-sphere" configuration are studied.
The unusual properties of the expectation values of the operators associated with the kinetic and potential energy of the two-electron atom/ion possessing the "electrons-on-sphere" configuration are found.
Refined formulas for calculations of the two-electron Fock expansion by the Green's function approach are presented.
The analytic wave functions of high accuracy describing the "electrons-on-sphere" configuration are obtained.
All results are illustrated in tables and figures.

\end{abstract}


\maketitle

\section{Introduction}\label{S0}

The properties of two interacting electrons confined to the surface of a sphere were always the subject of the intensitive investigations both experimental and theoretical.
As far back as almost 60 years ago the two electrons trapped in a harmonic external potential but repelling one another with the Coulomb interaction were studied \cite{KES}. Then the analytic solution for this system was obtained for a particular value of the harmonic force constant \cite{KAI} and, later, for a countably infinite set of force constants \cite{TAU}.
Related systems consisting of two electrons interacting through a Coulomb potential but confined within a
three-dimensional box with infinite walls \cite{ALA}, or ball of radius $r$ \cite{TH1,TH2}, were studied by the exact diagonalization technique.
The system of two electrons trapped on the surface of a sphere of radius $r$ has been used in Refs.\cite{EZR1,EZR2,OJH,HIN} to understand both weakly and strongly correlated systems and to suggest an "alternating" version of Hund`s rule \cite{WAR}.
In Ref.\cite{SEI1} the mentioned above systems were studied in the context of density-functional theory in order to test the interaction-strength interpolation model.
A comprehensive study of the singlet ground state of two electrons on the surface of a sphere of radius $r$ were performed in Ref.\cite{LO1}.

Note that the authors of all listed papers represented radius $r=|\textbf{r}_1|=|\textbf{r}_2|$ of sphere as a given parameter, and only the angle $\theta$ between the radius-vectors $\textbf{r}_1$ and $\textbf{r}_2$ of the electrons was considered as a quantum-mechanical variable.

In this paper we study the behavior of the two-electron atomic systems (another name is the helium-like isoelectronic sequence) in the S-state configuration which describes the situation when both electrons are located on the sphere of the radius $r$. We apply the Pekeris-like method (PLM) \cite{LEZ1,LEZ2} representing a fully three-body variational technique which consider both $r$ and $\theta$ as a quantum mechanical variables.

\section{The ground-state configuration of the helium-like atom/ion possessing both electrons located equidistantly from the nucleus} \label{S1}

We shall consider the S-state solution $\Psi\equiv\Psi(r_1,r_2,r_{12})$ of the non-relativistic Schrodinger equation
\begin{equation}\label{10}
H\Psi=E \Psi,
\end{equation}
where $E$ is the bound energy of the helium-like atom/ion with an infinitely massive nucleus of charge $Z$. The variables $r_1\equiv |\textbf{r}_1|$ and $r_2\equiv |\textbf{r}_2|$ represent the distances between each electron and the nucleus, whereas $r_{12}\equiv |\textbf{r}_1-\textbf{r}_2|$ is the inter-electron distance.
The Hamiltonian $H$ is defined, as usually, by the sum of the
kinetic energy operator
\begin{equation}\label{12}
T\equiv -\Delta/2,
\end{equation}
where $\Delta$ is the Laplacian, and
the potential energy operator $V$ representing the inter-particle Coulomb interactions
\begin{equation}\label{13}
V\equiv -\frac{Z}{r_1}-\frac{Z}{r_2}+\frac{1}{r_{12}}.
\end{equation}
The atomic units are used throughout this paper.

In our previous article \cite{LEZ3} the \emph{collinear} configuration
of the ground state of the two-electron atom/ion was studied.
The relevant \emph{collinear} wave function (WF) $\mathcal{K}(r,\lambda)\equiv\Psi(r,|\lambda|r,(1-\lambda)r)$ is the particular case of the general WF, $\Psi$ with the collinear parameter $\lambda\in[-1,1]$.
The relation that characterizes the \emph{collinear} arrangement of the particles is $\textbf{\emph{r}}_1=\lambda \textbf{r}_2$.

It is clear that the state when both electrons are situated on the surface of sphere of the radius $r$   is defined by the relation $r_1=r_2=r$.
For this case the inter-electron distance becomes $r_{12}=2r \sin(\theta/2)$  and the relevant WF is $\Phi(r,\theta)\equiv\Psi(r,r,2r\sin(\theta/2))$ with $\theta\in[0,\pi]$. Recall that the angle $\theta$ has been defined earlier as the angle between the radius-vectors of the electrons for the nucleus located at the origin.

We would like to emphasize the important features connecting both configurations of WF mentioned above.
The point is that the "\emph{collinear}" WF, $\mathcal{K}(r,\lambda)$ and the "\emph{electrons-on-sphere}" WF,  $\Phi(r,\theta)$ are coincident at the boundary values of their parameters $\lambda$ and $\theta$, respectively. In particular, for $\lambda=1\wedge \theta=0$ we obtain
\begin{equation}\label{14}
\mathcal{K}(r,1)=\Phi(r,0)=\Psi(r,r,0),
\end{equation}
which corresponds to the \emph{electron-electron} coalescence.
On the other hand, for $\lambda=-1\wedge \theta=\pi$ we obtain
\begin{equation}\label{15}
\mathcal{K}(r,-1)=\Phi(r,\pi)=\Psi(r,r,2r),
\end{equation}
which corresponds to the collinear configuration when the electrons are equidistantly on the opposite sides from the nucleus.

\begin{figure}
\caption{Schematic representation of the "collinear" and the "electrons-on-sphere" configurations.
The straight line (red online) corresponds to "collinear" configuration, whereas the semicircle (blue online) of the radius $r$ corresponds to the "electrons-on-sphere" configuration. The ball in the center of sphere represents the nucleus. $A$ and $B$ are the coupling points of both configurations.}
\includegraphics[width=6.0in]{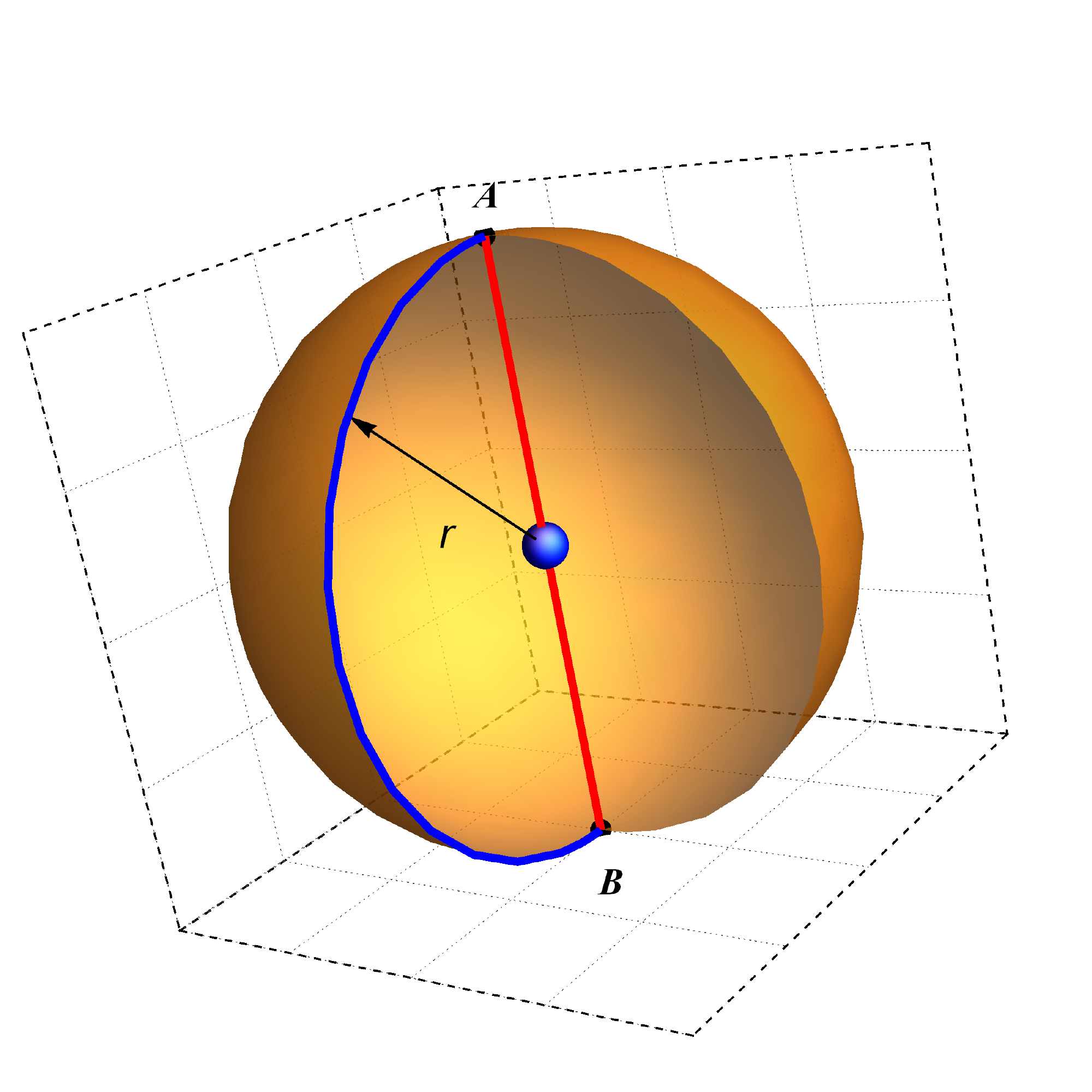}
\label{F1}
\end{figure}

The situation mentioned above is schematically shown in Fig.\ref{F1}.
It is seen that the "electrons-on-sphere" configuration covers the surface of sphere of the radius $r$, whereas the "collinear" configuration forms the diameter of this sphere.
Straight line (red online) crossing the nucleus  represents the "\emph{collinear}" configuration, whereas semicircle (blue online) of radius $r$ corresponds to the "\emph{electrons-on-sphere}" configuration.
Both curves are connected at the points $A$ and $B$.
Any two points
being corresponding to the positions $\textbf{r}_1$ and $\textbf{r}_2$ of the electrons define the inter-electron vector $\textbf{r}_{12}$.
Thus, any pair of points on the "\emph{collinear}" line corresponds to the definite $\lambda$ of the WF, $\mathcal{K}(r,\lambda)$. Accordingly, any pair of points on the "\emph{electrons-on-sphere}" semicircle corresponds to the definite angle $\theta$ (between the electrons) of the WF, $\Phi(r,\theta)$.
When one of the electrons is on the point $A$ and the second one is on the point $B$ (or vice versa), we obtain situation described by the WF (\ref{15}).
On the other hand, when both electrons are simultaneously located at the point $A$ or $B$, we obtain situation described by the WF (\ref{14}).

The basic expectation value characterizing the "\emph{collinear}" configuration is of the form \cite{LEZ3}:
\begin{equation}\label{16}
K(\lambda,Z)\equiv\left\langle\delta\left(\textbf{r}_1-\lambda \textbf{r}_2\right)\right\rangle=
4\pi \langle\delta(\textbf{r}_1)\delta(\textbf{r}_2)\rangle\int_0^\infty
\left|\mathcal{K}(r,\lambda)\right|^2r^2 dr,
\end{equation}
where $\mathcal{K}(0,\lambda)=1$, and $\delta(\textbf{r})$ is the three-dimensional delta function.
Note that expectation values $\langle\delta(\textbf{r}_1)\delta(\textbf{r}_2)\rangle$ being equal, in fact, to the square of the normalized WF taken at the nucleus, can be found in Ref. \cite{FR1,FR2,FR3} (see also references therein).

By analogy to the expectation value (\ref{16}),
for the WF, $\Phi(r,\theta)$ satisfying the boundary condition $\Phi(0,\theta)=1$,
we introduce the expectation value
\begin{equation}\label{17}
S(\theta,Z)=
4\pi \langle\delta(\textbf{r}_1)\delta(\textbf{r}_2)\rangle \mathcal{I}_Z(2,\theta)
\end{equation}
with
\begin{equation}\label{17a}
\mathcal{I}_Z(n,\theta)=
\int_0^\infty \left|\Phi(r,\theta)\right|^2r^n dr
\end{equation}
as the basic characteristic of the "electrons-on-sphere" configuration.
Both expectation values are certainly coincident for the boundary values of the parameters $\lambda$ and $\theta$ mentioned above (see points $A$ and $B$ in Fig.\ref{F1}), and for the given atom/ion, of course.

Using the PLM \cite{LEZ1,LEZ2} we have calculated the expectation values
$S(\theta,Z)$ for the ground states of the two-electron atomic systems with $1\leq Z\leq5$.
The values of $S(\theta,Z)$ with $0\leq\theta\leq \pi$ and step  $h_\theta=\pi/8$ are presented in Table \ref{T1}. The PLM parameter $\Omega$ (number of shells) for the given atom/ion was chosen under condition of the best coincidence with the published results of high accuracy (see, e.g., \cite{FR1,FR3,DRK}) for $S(0,Z)$ corresponding to the electron-electron coalescence.

\begin{figure}
\caption{The $\theta$-normalized expectation values $S(\theta,Z)$, given by Eq.\ (\ref{17}), as functions of the angle $\theta$ for the two-electron atomic systems considered. The solid line (black online) corresponds to the asymptotic two-electron ion with $Z\rightarrow \infty$.}
\includegraphics[width=6.0in]{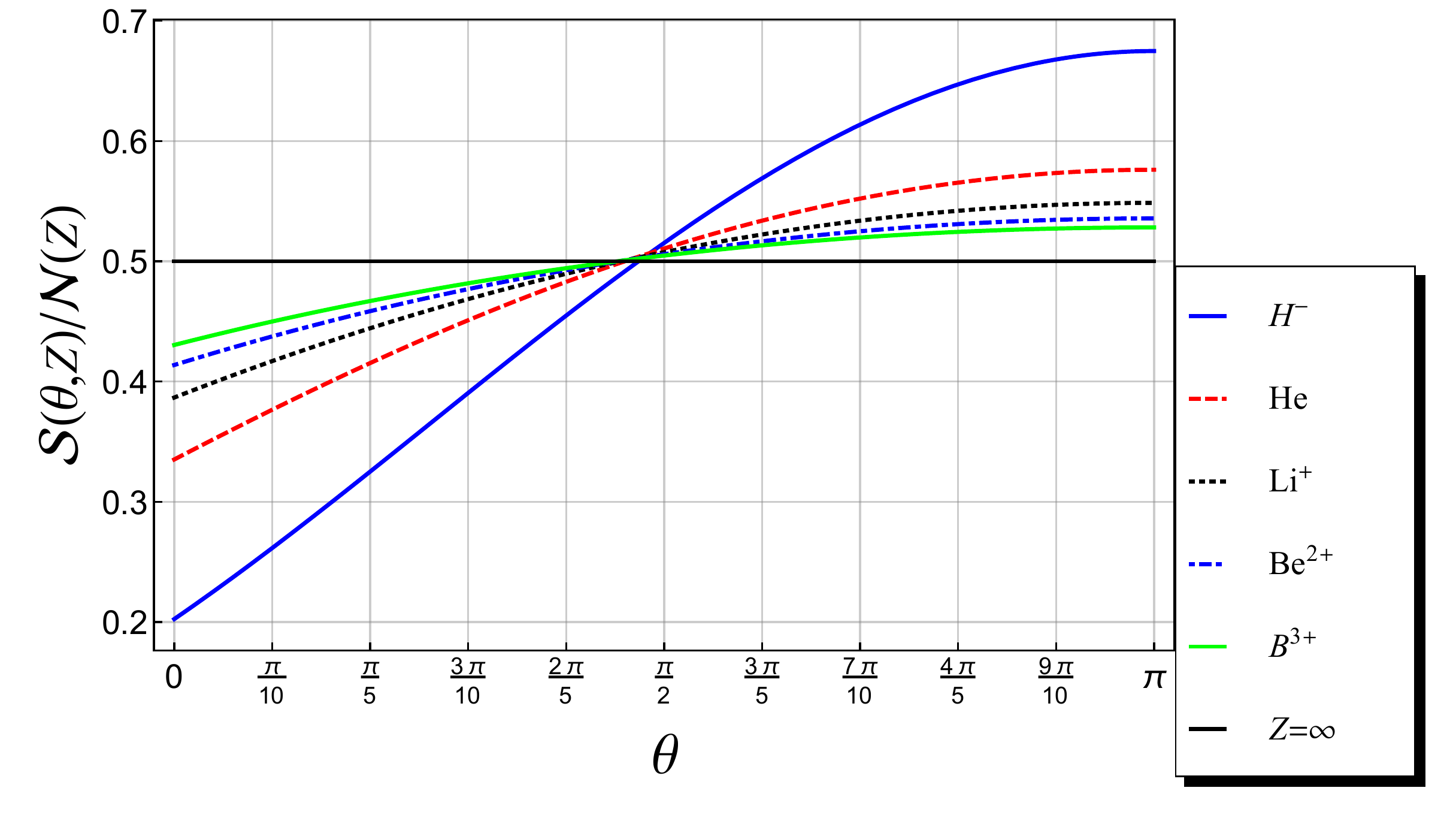}
\label{F2}
\end{figure}

Application of the normalization parameter over $\theta$, defined as
\begin{equation}\label{18}
\mathcal{N}(Z)=\int_{0}^\pi S(\theta,Z) \sin \theta  d\theta,
\end{equation}
enables us to place the plots of $\overline{S}_Z(\theta)\equiv S(\theta,Z)/\mathcal{N}(Z)$ for all considered $Z$ on a single figure (see Fig. \ \ref{F2}),
which demonstrates a rapid convergence of the curves with increasing $Z$.
To find the asymptotic curve ($Z\rightarrow \infty$),
let us suppose that for large enough $Z$ we can neglect the electron-electron interaction in comparison with the electron-nucleus interaction in the Schrodinger equation (\ref{10})-(\ref{13}).
It is well-known that the corresponding ground state solution is of the form $\Psi_{asymp}\sim\exp\left[-Z (r_1+r_2)\right]$ which for the case of the electrons on sphere reduces to
$\Phi_{asymp}(r,\theta)\sim\exp(-2 Z r)$.
Taking into account that $\langle\delta(\textbf{r}_1)\delta(\textbf{r}_2)\rangle=\Phi(0,\theta)/N$, where
\begin{equation}\label{19}
N=\int d^3\textbf{r}_2\int d^3\textbf{r}_1\left|\Psi(r_1,r_2,r_{12})\right|^2
\end{equation}
is the normalization integral, we obtain $N_{asymp}=\pi^2/Z^6$.
Subsequent substitution of $\Psi_{asymp}$ and $N_{asymp}$ into Eqs.(\ref{17}) and (\ref{18}) yields $S_{asymp}(\theta,Z)=Z^3/(8\pi)$ and $\mathcal{N}_{asymp}(Z)=Z^3/(4\pi)$, respectively,
resulting in the asymptotic expression
$\overline{S}_{\infty}(\theta)=\frac{1}{2}$.

The curves $\overline{S}_Z(\theta)$ (included the case of $Z\rightarrow \infty$) are shown in Fig.\ref{F2} for the helium atom, negative ion of hydrogen and for the positive two-electron ions with $Z=3,4,5$.
The interesting feature observed in Fig.\ref{F2} is the crossing of each curve by all others.
It is the most important to note that all of the mentioned intersections are located in the narrow range of angles. 
The right boundary of this range is $\theta_{max}\simeq 0.48555 \pi$ which corresponds to the intersection of the curves for H$^-$ $(Z=1)$ and He $(Z=2)$.
Note that the left boundary for the given ion/atom is defined by intersection of the corresponding curve with
the asymptotic curve $\overline{S}_{\infty}=1/2$.
In particular, for $\textrm{H}^-$-ion, we obtain $\theta_{min}\simeq 0.47384 \pi$.
To estimate the displacement of the left boundary, we have calculated (using the PLM) the expectation value $\overline{S}_{100}(\theta)$ corresponding to the two-electron positive ion with $Z=100$.
The corresponding curve crosses the asymptotic curve at the point
$\theta_{min}\simeq 0.44764 \pi$, which tells us about strong localization of the crossing points.

Due to the apparent proximity, all intersection points merge into one ravel in the Fig.\ref{F2}.

\section{Expectation values of the Hamiltonian for the WF with the "electrons-on-sphere" configuration} \label{S2}

It is clear that the Schrodinger equation (\ref{10}) must be satisfied for any configuration of the WF, $\Psi$.
Accordingly, let's do the following with this equation:
i) set $r_1=r_2=r$ and $r_{12}=2r\sin(\theta/2)$, ii) multiply on the left by $\Phi(r,\theta)r^2$,  iii) integrate over $r$ both sides of the resulting equation. This yields
\begin{equation}\label{30}
\int_0^\infty\Phi(r,\theta)\left[H \Psi(r_1,r_2,r_{12})\right]_{r_1=r_2=r}^{r_{12}=2r\sin(\theta/2)}r^2dr
=E \int_0^\infty\Phi^2(r,\theta)r^2 dr.
\end{equation}
Dividing both sides of Eq.(\ref{30}) by the RHS integral, we obtain the relation
\begin{equation}\label{31}
\langle T^{(\theta)}\rangle+\langle  V^{(\theta)}\rangle=E
\end{equation}
with the following notations.
The term $\langle T^{(\theta)}\rangle$ associated with the expectation value of the kinetic energy operator in the $S$-state "\emph{electrons-on-sphere}" configuration is of the form
\begin{equation}\label{32}
\langle T^{(\theta)}\rangle=
-\left[2~\mathcal{I}_Z(2,\theta)\right]^{-1}\int_0^\infty\Phi(r,\theta)
\left[\Delta \Psi(r_1,r_2,r_{12})\right]_{r_1=r_2=r}^{r_{12}=2r\sin(\theta/2)}r^2 dr,
\end{equation}
where the integral $\mathcal{I}_Z(2,\theta)$ is defined by Eq.(\ref{17a}), and the Laplacian is of the form (see e.g. \cite{GOT})
\begin{eqnarray}\label{33}
\Delta =\frac{\partial^2}{\partial r_1^2}+\frac{\partial^2}{\partial r_2^2}+2\frac{\partial^2}{\partial r_{12}^2}+
\frac{2}{r_1}\frac{\partial}{\partial r_1}+\frac{2}{r_2}\frac{\partial}{\partial r_2}+
\frac{4}{r_{12}}\frac{\partial}{\partial r_{12}}+
~\nonumber~~~~~~~~~~~~~~~~~~~~~~~~~~~~~~~~\\
\left(\frac{r_1^2-r_2^2+r_{12}^2}{r_1r_{12}}\right)\frac{\partial^2}{\partial r_1r_{12}}+
\left(\frac{r_2^2-r_1^2+r_{12}^2}{r_2r_{12}}\right)\frac{\partial^2}{\partial r_2r_{12}}.~~~~~~~~~~~~
\end{eqnarray}
Accordingly, for the term  $\langle V^{(\theta)}\rangle$ associated with the expectation value of the potential energy operator in the "\emph{electrons-on-sphere}" configuration, we obtain
\begin{equation}\label{34}
\langle V^{(\theta)}\rangle=
\frac{\mathcal{I}_Z(1,\theta)}{\mathcal{I}_Z(2,\theta)}\left(\frac{1}{2\sin(\theta/2)}-2Z\right).
\end{equation}
Note that
Eqs.(\ref{30})-(\ref{34}) are written for the real WFs because
we shall apply these equations to the PLM WFs which are actually real.
It is clear that Eqs.(\ref{30}) and (\ref{32}) can be easily transformed for the case of the complex WFs.

It follows from Eq.(\ref{31}) that functions $\langle T^{(\theta)}\rangle$ and $\langle V^{(\theta)}\rangle$ are symmetric in respect to the line $E/2$.
This means that the dimensionless functions $t_Z(\theta)=\langle T^{(\theta)}\rangle/|E|$ and $v_Z(\theta)=\langle V^{(\theta)}\rangle/|E|$ will be symmetric in respect to the line $(-1/2)$ which becomes the overall line of symmetry for all of the two-electron atoms.
Dividing the functions $\langle T^{(\theta)}\rangle$ and $\langle V^{(\theta)}\rangle$ by $|E|$ enables us also to preserve the signs and the zero positions for these functions.

It is seen from Eq.(\ref{34}) that a single zero $\theta_v$ of the function $v_Z(\theta)$ is: $\theta_v=2 \arcsin(1/4Z)$.
Accordingly, it follows from Eq.(\ref{31}) that a single zero $\theta_t$ of the function $t_Z(\theta)$ is represented by a root of equation $v_Z(\theta_t)=-1$.

\begin{figure}
\caption{Expectation values $t_Z(\theta)=\langle T^{(\theta)}\rangle/|E|$ (solid line with marker) and $v_Z(\theta)=\langle V^{(\theta)}\rangle/|E|$ (dashed line with marker) for the helium atom and all of the  two-electron ions under consideration. $\theta$ is the angle between the radius-vectors $\textbf{r}_1$ and $\textbf{r}_2$ of the electrons.
The extreme case of $Z=100$ is represented by solid line (black online) without marker.
The line of symmetry ($-1/2$) is presented (brown online).}
\includegraphics[width=5.5in]{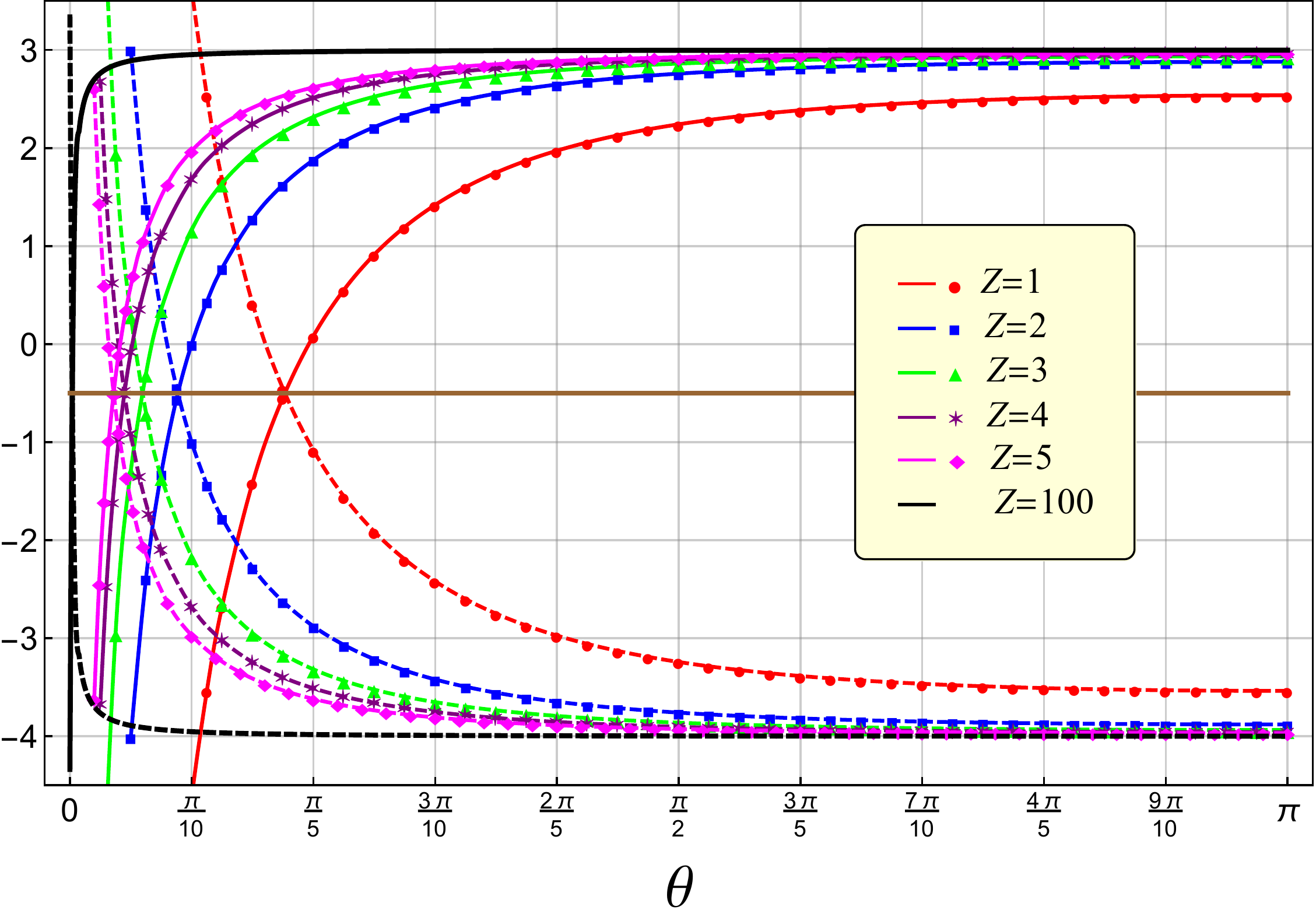}
\label{F3}
\end{figure}

The plots of the functions  $t_Z(\theta)$ and $v_Z(\theta)$ are presented in Fig. \ref{F3} for all of the helium-like atomic systems under consideration.
The line of symmetry $(-1/2)$ is displayed (in brown online) too.
To track the convergence of both functions with increasing $Z$, we have calculated (using PLM) the corresponding expectation values for very large nucleus charge $Z=100$ (see Fig. \ref{F3}).
The asymptotic case $Z\rightarrow \infty$  can be estimated as follows.
Let's suppose that for large enough $Z$ we can neglect the electron-electron interaction in comparison with the electron-nucleus interaction in the Schrodinger equation (\ref{10})-(\ref{13}).
It is well-known that the corresponding ground state solution is of the form $\Psi_{Z\rightarrow\infty}\sim \exp\left[-Z(r_1+r_2)\right]$. For the "electrons-on-sphere" configuration this yields $\Phi_{Z\rightarrow\infty}\sim \exp(-2 Z r)$. Using this WF, we can calculate (according to definition (\ref{17a})) the asymptotic integrals included into the representation (\ref{34}) for the expectation value $\langle V^{(\theta)}\rangle$. This yields:
\begin{equation*}
\mathcal{I}_{Z\rightarrow\infty}(1,\theta)=(16 Z^2)^{-1},~~~~~\mathcal{I}_{Z\rightarrow\infty}(2,\theta)=(32 Z^3)^{-1}.
\end{equation*}
Inserting these results into the RHS of Eq.(\ref{34}), we obtain for large enough $Z$:
\begin{equation*}
\langle V^{(\theta)}\rangle\underset{Z\rightarrow \infty}{=}
2Z\left(\frac{1}{2\sin(\theta/2)}-2Z\right).
\end{equation*}
Neglecting the electron-electron interaction, we obtain the mentioned above WF, $\Psi_{Z\rightarrow\infty}$ of two independent electrons with the well-known ground state energy equaled to $-Z^2/2$ per electron.
For the two-electron atomic system this yields $|E|\underset{Z\rightarrow \infty}{=}Z^2$.
Thus, for large enough $Z$ we obtain the dimensionless expectation values in the following analytic forms:
\begin{equation}\label{35}
v_Z(\theta)\underset{Z\rightarrow \infty}{=}\frac{1}{Z \sin(\theta/2)}-4,~~~~~
t_Z(\theta)\underset{Z\rightarrow \infty}{=}3-\frac{1}{Z \sin(\theta/2)}.
\end{equation}
Taking the limit of $v_Z(\theta)$ as $Z$ approaches infinity, one obtains $v_\infty(\theta\neq0)=-4$.
Using Eq.(\ref{31}) we accordingly obtain $t_\infty(\theta\neq0)=3$.
It is seen in Fig. \ref{F3} that our calculations (by the PLM) fully confirm these asymptotic estimates.
Note, that for $Z=100$ the analytic functions (\ref{35}) become visually indistinguishable from the corresponding functions calculated by the PLM.

Earlier we have described two characteristic angles $\theta_v$ and $\theta_t$ for which $v_Z(\theta_v)=0$ and $t_Z(\theta_t)=0$, respectively.
As it is seen from Fig. \ref{F3}  these angles are the boundary ones between which both expectation values
$\langle T^{(\theta)}\rangle$ and $\langle V^{(\theta)}\rangle$ are negative.
Moreover, there are two extra characteristic angles representing specific properties of the "electron-on-sphere" configuration.
The first one is the angle $\theta_{cr}$ of crossing the curves $\langle T^{(\theta)}\rangle$ and $\langle V^{(\theta)}\rangle$ (or $v_Z(\theta)$ and $t_Z(\theta)$, alternatively).
Using Eq.(\ref{31}) we can calculate this angle as the root of equation $v_Z(\theta_{cr})=-1/2$.
It was mentioned earlier that $\langle T^{(\theta)}\rangle$ and $\langle V^{(\theta)}\rangle$ are respectively associated with expectation values of the kinetic energy and potential energy operators in the "electrons-on-sphere" configuration.
Accordingly, the second of the extra characteristic angles is the angle $\theta_{vir}$ at which the mentioned expectation values obey the virial theorem for Coulomb interactions, that is $\langle V^{(\theta_{vir})}\rangle=-2\langle T^{(\theta_{vir})}\rangle$. Using Eq.(\ref{31}) one obtains the equivalent equation of the form $v_Z(\theta_{vir})=-2$.
All of the characteristic angles described above are presented in Table \ref{T2} for all of the two-electron atomic systems under consideration.

\section{The Fock expansion} \label{S3}

The behavior of the two-electron atomic WF, $\Psi(r_1,r_2,r_{12})$ near the nucleus is determined by the Fock expansion \cite{FOCK}
\begin{equation}\label{101}
\tilde{\Psi}(r_1,r_2,r_{12})\equiv\Psi(r_1,r_2,r_{12})/\Psi(0,0,0)=\sum_{k=0}^\infty R^k\sum_{p=0}^{[k/2]}\phi_{k,p}(\alpha,\theta)\ln^p R,
\end{equation}
where the hyperspherical coordinates $R,~\alpha\in [0,\pi]$ and $\theta\in [0,\pi]$ are defined as follows:
\begin{equation}\label{102}
R=\sqrt{r_1^2+r_2^2},~~~~\alpha=2\arctan\left(\frac{r_2}{r_1}\right),~~~~\theta=\arccos\left(\frac{r_1^2+r_2^2-r_{12}^2}{2 r_1 r_2}\right).
\end{equation}
It should be emphasized that the convergence of expansion (\ref{101}) had been proven in Ref. \cite{MORG1}.
Note that the hyperspherical angle $\theta$ coincides with the eponymous angle (between the radius-vectors of the electrons) introduced previously.
The explicit form of the angular Fock coefficients (AFC) $\phi_{k,p}(\alpha,\theta)$ for low orders $k$ can be found in Ref. \cite{LEZ4} (see also Refs. \cite{AB1, GOT}).
Clearly $\phi_{0,0}=1$ for the representation (\ref{101}).

For the "electrons-on-sphere" configuration when $R=r \sqrt{2}$ and $\alpha=\pi/2$, the Fock expansion (\ref{101}) becomes:
\begin{equation}\label{103}
\Phi(r,\theta)\underset{r\rightarrow 0}{=}1+c_1 r+c_{2L} r^2 \ln r+c_2 r^2+c_{3L} r^3\ln r+c_3 r^3+O(r^4).~~
\end{equation}
It follows from the general expansion (\ref{101}) that all $c-$coefficients of expansion (\ref{103}) can be expressed in terms of the AFCs and/or its components $\phi_{k,p}^{(j)}(\pi/2,\theta)$.
It is worth noting that calculation of the AFCs is a complicated problem.
The most (but not all!) of the AFC-components for $k\leq 4$ have been derived in the explicit (analytic) form \cite{LEZ4} (see also \cite{LEZ5,LEZ6} for $k \geq 4$).
Substantial success in solving the problem can be achieved by the Green's function (GF) approach.

The possibility of calculations of the AFCs by the GF method  has been declared still in the original paper of Fock \cite{FOCK}. Clarification and concretization of the results reported in this work lead to the following formulas for the AFC-component calculations:
\begin{equation}\label{104}
\phi_{k,p}^{(j)}(\alpha',\theta')=\frac{1}{8\pi}\int_0^\pi d\alpha \sin^2\alpha\int_0^\pi d\theta \sin \theta~ h_{k,p}^{(j)}(\alpha,\theta)
\int_0^\pi \frac{\cos\left[\left(\frac{k}{2}+1\right)\omega \right]}{\sin \omega}d\varphi,~~~~~~~~[k~~odd]
\end{equation}
\begin{equation}\label{105}
\phi_{k,p}^{(j)}(\alpha',\theta')=\frac{1}{8\pi^2}\int_0^\pi d\alpha \sin^2\alpha\int_0^\pi d\theta \sin \theta~ h_{k,p}^{(j)}(\alpha,\theta)
\int_0^\pi \frac{\cos\left[\left(\frac{k}{2}+1\right)\omega \right]}{\sin \omega}(\pi-\omega)d\varphi,~~~~[k~~even]
\end{equation}
where $\omega \in [0,\pi]$ denotes an angle defined by the relation \cite{FOCK}
\begin{equation}\label{106}
\cos \omega =\cos \alpha \cos \alpha'+\sin \alpha \sin \alpha'
\left[\cos \theta \cos \theta'+\sin \theta \sin \theta' \cos(\varphi-\varphi')\right]
\end{equation}
with auxiliary angle $\varphi \in [0,2\pi]$. The RHS, $h_{k,p}^{(j)}$ of the individual Fock recurrence relation (IFRR)
\begin{equation}\label{107}
\left[ \Lambda^2-k(k+4)\right]\phi_{k,p}^{(j)}\left(\alpha,\theta\right)=h_{k,p}^{(j)}\left(\alpha,\theta\right)
\end{equation}
has been defined, e.g., in Refs. \cite{LEZ4} or \cite{LEZ7}, whereas
\begin{equation}\label{108}
\Lambda^2=-4\left[\frac{\partial^2}{\partial\alpha^2}+2\cot\alpha\frac{\partial}{\partial\alpha}+ \frac{1}{\sin^2\alpha}\left( \frac{\partial^2}{\partial \theta^2}+\cot \theta \frac{\partial}{\partial\theta}\right)\right]
\end{equation}
is the hyperspherical angular momentum operator projected on $S$ states.

Remind the connection between the AFC, $\phi_{k,p}(\alpha,\theta)$ and the AFC-components $\phi_{k,p}^{(j)}(\alpha,\theta)$, as well as between the RHS of the corresponding Fock recurrence relation (FRR) and IFRR:
\begin{equation}\label{109}
\phi_{k,p}(\alpha,\theta) =\sum_{j=p}^{k-p}Z^j\phi_{k,p}^{(j)}\left(\alpha,\theta\right),~~~~~~~
h_{k,p}(\alpha,\theta) =\sum_{j=p}^{k-p}Z^jh_{k,p}^{(j)}\left(\alpha,\theta\right).
 \end{equation}
It is worth noting that the GF formulas (\ref{105})-(\ref{106}) enable us to calculate only  the so called "pure" AFC-components, $\phi_{k,p}^{(j)}(\alpha,\theta)$ \cite{LEZ4} (for even $k$) which do not contain the admixture of the hyperspherical harmonics (HH) $Y_{kl}(\alpha,\theta)$ satisfying the homogeneous differential equation associated with the IFRR (\ref{107}).

We verified the validity of the GF formulas (\ref{104})-(\ref{106}) on examples of all AFCs known to us.
We have found that representation (\ref{105}) for even values of $k$ is correct only for angle $\varphi'=0$ or $\varphi'=\pi$, unless representation (\ref{104}) for odd $k$ that is correct for any $\varphi'$. Thus, we believe that using of the angle $\varphi'=0$ represents the general case which is the most simple one, as well.
Note that the particular case $\theta'=0$ was considered in Ref. \cite{LEZ7}. It follows from Eq.(\ref{106}) that for this case angle $\omega$ is independent on the angle $\varphi$. Whence, integration over $\varphi$ in (\ref{104}) or (\ref{105}) yields $\pi$, and we obtain the GF formulas presented in Ref. \cite{LEZ7}.

Using the AFC-components derived in Refs.\cite{LEZ4,LEZ5,LEZ6,LEZ7} we have calculated the coefficients $c_1,c_{2L},c_{3L}$  and $c_2$ (of the Fock expansion (\ref{103})) in the explicit form as functions of the angle $\theta$ (see the Appendix).
The most of the AFC-components associated with calculation of the coefficient $c_3$ can be obtained by the method described in Ref.\cite{LEZ4}. However, the AFC-subcomponents $\chi_{30}(\pi/2,\theta), \chi_{31}(\pi/2,\theta)$ and $\phi_{3,0}^{(2e)}(\pi/2,\theta)$ (see the Appendix) can be calculated only numerically by the GF approach described above. The corresponding results of high accuracy for $\theta=0$ up to $\theta=\pi$ with step $h_\theta=\pi/8$ are presented in Table \ref{T3}.
The details of all calculations can be found in the Appendix.
One should emphasize that calculations of the coefficients $c_1,c_{2L}$ and $c_{3L}$ are dependent on two parameters only which are well-known.
These are the nucleus charge $Z$ and the non-relativistic energy $E$  of the two-electron atom/ion (see, e.g., \cite{DRK}).
As to the coefficients  $c_2$ and $c_3$ then it is important to note that the corresponding calculations include the extra parameter $a_{21}$ (see the Appendix) which has not been reliably calculated previously (see also \cite{LEZ3}).

\section{Analytic wave functions of high accuracy} \label{S4}

In this Section, we propose two methods for obtaining
the analytic WFs (of high accuracy) describing the "electron-on-sphere" configuration of the two-electron atom/ion.

The Schrodinger equation (\ref{10}) expressed in the hyperspherical coordinate (\ref{102}) can be written in the form
\begin{equation}\label{111}
\frac{\partial^2 \tilde{\Psi}}{\partial R^2}+\frac{5}{R}\frac{\partial \tilde{\Psi}}{\partial R}-\frac{1}{R^2}\Lambda^2\tilde{\Psi}+
2\left[\frac{1}{R}\left(Z V_1-V_0\right)+E
\right]\tilde{\Psi}=0,
\end{equation}
where $\tilde{\Psi}\equiv\tilde{\Psi}(R,\alpha,\theta)=\Psi(r_1,r_2,r_{12})$, whereas
\begin{equation}\label{112}
V_0=\frac{1}{\sqrt{1-\sin \alpha\cos\theta}},~~~~~~V_1=\frac{2\sqrt{1+\sin \alpha}}{\sin\alpha}
\end{equation}
are the angular parts of the inter-particle potential (\ref{13}), $R V\equiv V_0-Z V_1$. The operator $\Lambda^2$ is defined by Eq.(\ref{108}).
For the "electrons-on-sphere" configuration (when $R=r\sqrt{2}$ and $\alpha=\pi/2$) the Schrodinger equation (\ref{111}) becomes
\begin{equation}\label{113}
\frac{\partial^2 \Phi}{\partial r^2}+\frac{5}{r}\frac{\partial \Phi}{\partial r}+\frac{4}{r^2}
\left[X+\frac{\partial^2\Phi}{\partial \theta^2}+\cot\theta\frac{\partial\Phi}{\partial \theta}\right]+
4\left[\frac{1}{r}\left(2Z-\frac{1}{2\sin(\theta/2)}\right)+E\right]\Phi=0,
\end{equation}
where the WF, $\Phi\equiv\Phi(r,\theta)=\tilde{\Psi}(r\sqrt{2},\pi/2,\theta)$ was introduced in Sec.\ref{S1}, and
\begin{equation}\label{114}
X\equiv X(r,\theta)=\left.\frac{\partial^2\tilde{\Psi}(r\sqrt{2},\alpha,\theta)}{\partial\alpha^2}\right|_{\alpha=\pi/2}.
\end{equation}

The PLM calculations for the ground state (at least) of the helium-like atom/ion show the high accuracy of approximation
\begin{equation}\label{115}
X+\frac{\partial^2\Phi}{\partial \theta^2}+\cot\theta\frac{\partial\Phi}{\partial \theta}\simeq
r\left[(a+b r)\Phi+(c+d r)\frac{\partial\Phi}{\partial r}\right],
\end{equation}
where $a,b,c$ and $d$ are  the parameters.
Substitution of the approximation (\ref{115}) into the Schrodinger equation (\ref{113}) enables us
to believe that the WF, $F(r)\equiv\Phi(r,\theta_0)$ at the given angle $\theta_0\in[0,\pi]$ satisfies the equation
\begin{equation}\label{116}
F''(r)+\left(\frac{A}{r}+B\right)F'(r)+\left(\frac{C}{r}+D\right)F(r)=0,
\end{equation}
where the parameters $A,B,C$ and $D$ are currently undetermined.
The general solution  of Eq.(\ref{116}) is:
\begin{equation}\label{117}
F(r)=e^{-\frac{1}{2}(B+\sigma)r}\left[\bar{c}_1 U(\kappa,A,\sigma r)+\bar{c}_2 L_{-\kappa}^{(A-1)}(\sigma r)\right],
\end{equation}
where
\begin{equation}\label{118}
\sigma=\sqrt{B^2-4D},~~~~~~~~~~~~\kappa=\frac{A}{2}+\frac{AB-2C}{2\sigma}.
\end{equation}
Considering the behavior of the special functions at the origin ($r\rightarrow 0$), one can conclude that the series expansion of the generalized Laguerre function $L_{-\kappa}^{(A-1)}(\sigma r)$ does not contain terms with $\ln r$, whereas the Tricomi confluent hypergeometric function $U(\kappa,A,\sigma r)$ does contain logarithmic terms of the form $r^{n+1} \ln r$, but only if the parameter $A=-n$, where $n$ is the positive integer.
Such properties of the relevant Tricomi function are similar to those of the Fock expansion (\ref{103}) if the additional condition $n=1$ is imposed (there is only one exception which will be discussed later).
For $A=-n$ the general solution of Eq.(\ref{116}) is of a special form
\begin{equation}\label{119}
F_n(r)=e^{-b r}r^{n+1}\left[\bar{c}_1 U(\kappa_n,n+2,\sigma r)+\bar{c}_2 L_{-\kappa_n}^{(n+1)}(\sigma r)\right],
\end{equation}
where we denoted
\begin{equation}\label{120}
b=\frac{1}{2}(B+\sigma),~~~~~~~~~~~~\kappa_n=1+\frac{n}{2}-\frac{2C+nB}{2\sigma},
\end{equation}
whereas the parameter $\sigma$ is defined by Eq.(\ref{118}).
In its turn, it can be shown that the asymptotic behavior ($r\rightarrow \infty$) of the function $L_{-\kappa_n}^{(n+1)}(\sigma r)$ is characterized by the exponential $\exp(\sigma r)$ which is divergent for $\sigma>0$ (see Eq.(\ref{118})).
This implies that the WF (\ref{119}) tends asymptotically to zero only under condition $\textrm{Re}(B)>\textrm{Re}(\sigma)$.

Given the argumentation mentioned above we shall consider two options for constructing the model WF of high accuracy which describes the "electrons-on-sphere" configuration.

\subsection{Single-term model WF }\label{S4a}

First, let's set $\bar{c}_2=0$ and $n=1$ in the general solution (\ref{119}) to build the simplest model WF of the form
\begin{equation}\label{122}
F(r)=e^{-b r}(\sigma r)^2 \Gamma(\kappa)U(\kappa,3,\sigma r),
\end{equation}
which satisfies the condition $F(0)=1$. Here $\Gamma(\kappa)$ is the Euler gamma function, whereas $\kappa\equiv\kappa_1$ according to definition (\ref{120}).
It is seen that the model WF (\ref{122}) contains 3 parameters $\kappa, \sigma$ and $b$, which calculations require three (at least) coupling equations (CE). The first CE can be obtained by equating the coefficients for $r$ in the series expansion of the WF (\ref{122}) and the Fock expansion (\ref{103}). This yields
\begin{equation}\label{123}
b+\sigma(\kappa-2)+c_1=0,
\end{equation}
where the coefficient $c_1$ is determined in the Appendix.

Equating successively the coefficients for $r^2$ and $r^3$ in the series expansion of the WF (\ref{122}) and the Fock expansion (\ref{103}), one obtains the equations
\begin{equation}\label{124}
2b^2+4b\sigma(\kappa-2)+\sigma^2(\kappa-2)(\kappa-1)(3-2\boldsymbol\varsigma)=4c_2,
\end{equation}
\begin{equation}\label{125}
\sigma^2(\kappa-2)(\kappa-1)\left[\sigma(17\kappa-6)-27b+6\boldsymbol\varsigma(3b-\kappa \sigma)\right]
-18b^2 \sigma(\kappa-2)-6b^3=36 c_3,
\end{equation}
where the auxiliary identifier
\begin{equation}\label{126}
\boldsymbol\varsigma\equiv\ln \sigma+ 2\gamma+\psi_0(\kappa)
\end{equation}
includes the Euler constant $\gamma$ and the digamma function $\psi_0(\kappa)$.
Calculations of the coefficients $c_2$ and $c_3$ of the Fock expansion (\ref{103}) are described in the Appendix in details. The problem is that the calculation formulas for these coefficients contain the parameter $a_{21}$ characterizing the contribution of the HH, $Y_{21}(\alpha,\theta)$ into the AFC, $\phi_{2,0}(\alpha,\theta)$ (see, e.g., \cite{LEZ4}, \cite{MYERS}).
To date, there are no reliable calculations of the parameter $a_{21}$.
Fortunately, both coefficient $c_2$ and $c_3$ are linearly dependent on this parameter which enables us to eliminate it between the set of Eqs.(\ref{124}) and (\ref{125}).
The result is the second (transcendental) CE for the parameters $\kappa, \sigma$ and $b$.
At last, we propose to use the equation
\begin{equation}\label{127}
4\pi \langle\delta(\textbf{r}_1)\delta(\textbf{r}_2)\rangle
\int_0^\infty |F(r)|^2 r^2 dr=S(\theta,Z),
\end{equation}
as the third CE we are looking for.
Function $F(r)$ represented in Eq.(\ref{127}) is the single-term model WF of the form (\ref{122}).
The expectation values $S(\theta,Z)$ were discussed in Sec. \ref{S1} (see Eq.(\ref{17})).
Note that the values of $S(0,Z)\equiv\langle\delta(\textbf{r}_1-\textbf{r}_2)\rangle $ corresponding to the specific case of the electron-electron coalescence can be found, e.g., in Refs. \cite{DRK}, \cite{FR3}. The values of $S(\pi,Z)\equiv\langle\delta(\textbf{r}_1+\textbf{r}_2)\rangle $
corresponding to the specific case of the collinear $\textbf{e-n-e}$ configuration
have been published in Ref.\cite{LEZ3}.
The intermediate expectation values $S(\theta,Z)$ calculated by the PLM (see also Sec. \ref{S2}) are presented in Table \ref{T1}.

The first CE (\ref{123}) enables us to express any of 3 parameters $\kappa, \sigma$ and $b$ in terms of two another ones. Inserting the resulting relation into the second and third CEs, we obtain the set of two transcendental equations for two of 3 parameters we are looking for.
These set of equations can be solved, for example, by the Wolfram Mathematica built-in program $\textbf{FindRoot}$.
Parameters of the model WF (\ref{122}) for helium ($Z=2$) are presented in Table \ref{T4}, as an example of the technique described above.
These parameters are shown for different cases of the mutual arrangement of electrons (characterizing by the angle $\theta$) on the sphere of the radius $r$.

These results require some important comments.
First, it follows from Table \ref{T4} that for some values of $\theta$ the parameters can be complex.
Second, to build the model WF (\ref{122}) we have selected $n=1$ in the general solution (\ref{119}), because the logarithmic series of the Fock expansion starts (in general) with the term $r^2\ln r$.
There is only one exception when such series starts with the term $r^3\ln r$. This is the case of $\theta=\pi/2$ for the "electrons-on-sphere" configuration. It is clear that for this specific case one should either select $n=2$ or set to zero the coefficient, $-(\kappa-2)(\kappa-1)\sigma^2/2$ for the $r^2\ln r$ in the series expansion of the WF of the form (\ref{122}).
It is clear that selecting the second option one should set $\kappa=2$ or $\kappa=1$. Parameters $b$ and $\sigma$ for $\kappa=1$ are presented in Table \ref{T4}.
The last comment is related to estimation of the accuracy of the model WF for a given $\theta$ presented in each separate row of the Table \ref{T4}.
As a measure of this accuracy, we chose the value
\begin{equation}\label{128}
\mathcal{R}=\int_0^\infty r|F(r)-\Phi(r,\theta)|dr\left(\int_0^\infty r \Phi(r,\theta)dr\right)^{-1}
\end{equation}
represented in the right column of the Table \ref{T4}. Here $F(r)$ is the model WF (\ref{122}) and $\Phi(r,\theta)$ is an actual WF calculated by the PLM.

\subsection{Double-term model WFs }\label{S4a}

It can be shown that parameter $\kappa_n$ defined by Eq.(\ref{120}) satisfies the relation
\begin{equation}\label{130}
\kappa_n=\kappa+(n-1)\left(1-\frac{b}{\sigma}\right),
\end{equation}
where $\kappa\equiv\kappa_1$ corresponds to the initial case with $n=1$.

The use of Eq.(\ref{130}) enables us
to consider the model WF of the form:
\begin{equation}\label{131}
F(r)=e^{-b r}(\sigma r)^2\left[\lambda \Gamma(\kappa)U(\kappa,3,\sigma r)+
\left(\frac{1-\lambda}{2}\right) \Gamma\left(\kappa+1-\frac{b}{\sigma}\right)\sigma r
U\left(\kappa+1-\frac{b}{\sigma},4,\sigma r\right)\right].
\end{equation}
The WF (\ref{131}) represents the linear combination of the Tricomi functions containing in the functions $F_n(r)$ defined by Eq.(\ref{119}).
In accordance to Eq.(\ref{130}) for $\kappa_n$, we set $n=1$ for the first Tricomi function and $n=2$ for the second one.
The second Tricomi function describes the effect of the electron-electron coalescence (for the angles $\theta$ close to $\pi/2$) has been mentioned before.
The coefficients are chosen such a way that $F(0)=1$.
Note that in addition to the variable $r$, the model WF (\ref{131}) depends on 4 parameters $b,\sigma,\kappa$ and $\lambda$. The latter parameter characterizes contribution of each of both Tricomi functions.

Unlike the single-term calculations,
the double-term ones are based solely on the Fock expansion.
Consequently, and it is important to emphasize, this version of calculations requires knowledge of only two physical parameters: nucleus charge $Z$ and the non-relativistic electron energy $E$ of the two-electron atom/ion.

Equating successively the coefficients for $r,~r^2\ln r,~r^3\ln r,~r^2$ and $r^3$ in the series expansion of the WF (\ref{131}) and the Fock expansion (\ref{103}), one obtains five CEs of the form:
\begin{equation}\label{132}
(\lambda+1)\left[b+\sigma(\kappa-2)\right]=-2c_1,
\end{equation}
\begin{equation}\label{133}
\lambda \sigma^2(\kappa-2)(\kappa-1)=-2c_{2L},
\end{equation}
\begin{eqnarray}\label{134}
b^2(\lambda-1)\left[b-3 \sigma (\kappa-1)\right]+
b\sigma^2\left\{14\lambda-2+3\kappa\left[2-8\lambda+\kappa(3\lambda-1)\right]
\right\}
~~~~~\nonumber~~~~~~\\
-\sigma^3\kappa(\kappa-1)(\kappa-2)(3\lambda-1)=12c_{3L},
\end{eqnarray}
\begin{equation}\label{135}
b^2(\lambda+1)+b\sigma (4\kappa\lambda-7\lambda-1)+\sigma^2(\kappa-2)(\kappa-1)\left[2\lambda(1
-\boldsymbol\varsigma)+1\right]=4c_2,
\end{equation}
\begin{eqnarray}\label{136}
\sigma\Big(3b^2\left[17\lambda+7-\kappa(7\lambda+5)\right]+b\sigma\left\{3\kappa\left[\kappa(5-23\lambda)+
62\lambda-8\right]
-112\lambda+4\right\}
~~~~~~~~~~~~~~~~\nonumber~~\\
\left.
+\sigma^2(\kappa-2)(\kappa-1)\left[9\lambda(5\kappa-2)-11\kappa+6\right]
\Big)+
6(\ln \sigma+2\gamma)\Big(b^3(\lambda-1)-3b^2\sigma(\kappa-1)(\lambda-1)
\right.
~~~~~~~~\nonumber~~\\.
+b \sigma^2
\left\{3\kappa\left[\kappa(3\lambda-1)-8\lambda+2\right]+14\lambda-2\right\}
-\sigma^3\kappa(\kappa-1)(\kappa-2)(3\lambda-1)
\Big)
~~~~~~~~~~~~~~~~~~\nonumber~~\\
-b^3(11\lambda+1)-12\lambda \sigma^2(\kappa-1)(\kappa-2)(\kappa\sigma-3b)\psi_0(\kappa)
~~~~~~~~~~~~~~~~~~\nonumber~~~~~~\\
+6(\lambda-1)\left[b-\sigma(\kappa-2)\right](b-\kappa \sigma)
[b-\sigma(\kappa-1)]\psi_0\left(\kappa-\frac{b}{\sigma}\right)=72 c_3.~~~~~~~~~~~
\end{eqnarray}
Note that calculation of the coefficients $c_1,c_{2L},c_{3L},c_2$ and $c_3$ is described in the Appendix.
Fortunately, the set of 3 equations (\ref{132}), (\ref{133}) and (\ref{134}) can be solved analytically in respect to the parameters $\kappa,\sigma$ and $b$ (in terms of the parameter $\lambda$). This gives us four solutions of the form:
\begin{equation}\label{137}
b=\frac{c_1 \lambda(\lambda-1)[\varpi(\lambda+1)-5c_1]-\rho-2c_{2L}(2\lambda-1)(\lambda+1)^2}{4c_1 \lambda(\lambda^2-1)},
\end{equation}
\begin{equation}\label{138}
\sigma=\frac{\varpi}{2},~~~~~~~\kappa=\frac{3}{2}-\frac{\varpi\lambda(\lambda+1)(\rho x_2+x_1)}{8x_3},
\end{equation}
where the auxiliary identifiers are
\begin{equation}\label{139}
x_1=\left[2c_1^3\lambda(\lambda-1)+3c_{3L}\lambda(\lambda+1)^3-2c_1c_{2L}(\lambda+2)(\lambda+1)^2\right]
\left[3c_1^2\lambda(\lambda-1)-2c_{2L}(2\lambda-1)(\lambda+1)^2\right],
\end{equation}
\begin{equation}\label{140}
x_2=2c_1^3\lambda(\lambda-1)+3c_{3L}\lambda(\lambda+1)^3-6c_1c_{2L}(3\lambda-2)(\lambda+1)^2,
\end{equation}
\begin{eqnarray}\label{141}
x_3=32c_{2L}^3(2\lambda-1)^2(\lambda+1)^6+\lambda^3\left[2c_1^3(\lambda-1)+3c_{3L}(\lambda+1)^3\right]^2
~\nonumber~~~~~~~~~~~~~~~~~~~~~\\
\left.
-12c_1c_{2L}\lambda^2(\lambda+1)^2\left[2c_1^3(\lambda-1)+3c_{3L}(3\lambda-2)(\lambda+1)^3\right]
\right.
+12c_1^2c_{2L}^2\lambda(\lambda+1)^4\left[4+\lambda(11\lambda-12)\right],
\nonumber\\~~~
\end{eqnarray}
\begin{eqnarray}\label{142}
\rho=
\pm\sqrt{4c_{2L}^2(2\lambda-1)^2(\lambda+1)^4+c_1\lambda(\lambda-1)
\left[c_1^3\lambda(\lambda-1)-12c_{3L}\lambda(\lambda+1)^3+4c_1c_{2L}(4\lambda-1)(\lambda+1)^2\right]},
\nonumber\\~~~
\end{eqnarray}
\begin{eqnarray}\label{143}
\varpi=\pm\left\{8c_{2L}^2(2\lambda-1)^2(\lambda+1)^4-2c_1\lambda(\lambda-1)
\left[3 c_1 \rho-5c_1^3\lambda(\lambda-1)+6c_{3L}\lambda(\lambda+1)^3\right]
\right.
\nonumber\\~~~
\left.
+4c_{2L}(\lambda+1)^2\left[6c_1^2\lambda(\lambda-1)^2+(2\lambda-1)\rho \right]
\right\}^{\frac{1}{2}}/\left[c_1 \lambda(\lambda^2-1)\right].~~~~~~~~~~~
\end{eqnarray}
Note that combinations of the different signs for $\rho$ and $\varpi$ produce four different solutions.
However, it can be verified that only one solution with both positive signs (mentioned above) reproduces the physical situation.

Similar to how it was done in the previous section, we need to eliminate the parameter $a_{21}$ (containing in $c_2$ and $c_3$) between Eqs.(\ref{135}) and (\ref{136}) to obtain a single transcendental equation including the parameters $b,\sigma,\kappa$ and $\lambda$. The subsequent substitution of the representations (\ref{137})-(\ref{143}) for $b,\sigma$ and $\kappa$ into the resulting equation transforms it into the complicated transcendental equation of only one parameter $\lambda$. Note that the corresponding solutions of this equation for $\lambda$, as well as the other parameters $b,\sigma$ and $\kappa$ calculated by Eqs. (\ref{137})-(\ref{143}) can be complex.
Parameters of the model WF (\ref{131}) of helium are presented in Table \ref{T5} for different values of $\theta$.
It is seen that for $\theta=0,\pi/8,\pi/4$ (when the configuration of WF is close to the electron-electron coalescence) all parameters are complex. Note that one obtains the same $|F(r)|$ if to provide the complex conjugation of all (four) of the corresponding parameters simultaneously.

\newpage

\section{Conclusions} \label{S5}

The properties of the "electrons-on-sphere" configuration of the helium atom and the two-electron ions have been studied.
The corresponding wave function describes the special quantum-mechanical state of the atomic system when both electrons are located on the surface of a sphere of the radius $r$, and the angle $\theta$ characterizes the mutual arrangement of the electrons on sphere.
Unlike the previous studies
we considered $r$ as a quantum mechanical variable but not as a parameter.
It is worth noting that
the "electrons-on-sphere" and the "collinear" configuration \cite{LEZ3} are coincident in two boundary points.
For $\theta=0$ one obtains the state of the electron-electron coalescence, whereas the angle $\theta=\pi$ characterizes
the $\textbf{e-n-e}$ configuration when the electrons are located at the ends of the diameter of sphere with the nucleus at its center (see Fig. \ref{F1}).

By analogy to the expectation value (\ref{16}) representing  the "collinear" configuration,
we have introduced the expectation value $S(\theta,Z)$ characterizing the "electrons-on-sphere" configuration (see Eqs.(\ref{17})-(\ref{17a})).
Using the Pekeris-like method \cite{LEZ1,LEZ2} we have calculated the expectation values
$S(\theta,Z)$ for the ground states of the two-electron atomic systems with $1\leq Z\leq5$.
The results are presented in Table \ref{T1}. A strong localization of the intersection points of the curves corresponding to different $Z$ (atom/ions) has been revealed in a narrow range around $\theta\simeq 0.47 \pi$ (see Fig. \ref{F2}).

The expectation values of the dimensionless operators $v_Z(\theta)$ and $t_Z(\theta)$ associated with the potential and kinetic energy, respectively, of the two-electron atom/ion possessing the "electrons-on-sphere" configuration were calculated.
The characteristic angles describing the unusual properties of these expectation values were found (see Table \ref{T2}).
For example, we have calculated the angles $\theta_v$ and $\theta_t$ between which both $v_Z(\theta)$ and $t_Z(\theta)$ are negative (see Fig. \ref{F3}).

Refined formulas (\ref{104})-(\ref{106}) for calculation of the angular Fock coefficients by the Green's function approach
were presented. These (GF) approach enabled us to calculate numerically some AFC-components that cannot be calculated by another methods (see the Appendix).

The analytic WFs of high accuracy were derived for the ground state of the two-electron atom/ion possessing the "electron-on-sphere" configuration.
The first kind of the model (single-term) WF  was build as the product of the Tricomi confluent hypergeometric function and exponential (see Eq.(\ref{122})). Calculation of its parameters was based on both the Fock expansion (see the Appendix) and the characteristic expectation values (\ref{17}). The corresponding parameters for helium can be found in Table \ref{T4}.
The second kind of the model (double-term) WF was represented by the product of the linear combination of two Tricomi functions and
exponential (see Eq.(\ref{131})). It is worth noting that the parameters of this WF can be calculated by the use of the Fock expansion exclusively and nothing more.
This implies that the input data for the relevant calculations are represented by two physical parameters only, the non-relativistic electron energy $E$ and the nucleus charge $Z$ of the two-electron atom/ion.
The corresponding parameters for helium can be found in Table \ref{T5}.
The results presented in the last two tables show that for some angles $\theta$ the parameters of both model WFs can be complex quantities.

\section{Acknowledgment}

This work was supported by the PAZY Foundation.
We acknowledge helpful discussions with Prof. N. Barnea and Prof. R. Krivec.

\appendix

\section{}\label{SA}

It follows from the general expansion (\ref{101}) that the coefficients of the particular Fock expansion (\ref{103}) for the "electrons on sphere" configuration can be calculated by the formulas:
\begin{equation}\label{A1}
c_1=\sqrt{2}\left[\phi_{1,0}^{(0)}+Z\phi_{1,0}^{(1)}\right]=\sin(\theta/2)-2Z,
\end{equation}
\begin{equation}\label{A2}
c_{2L}=2Z\phi_{2,1}^{(1)}=-\frac{2 Z(\pi-2)}{3\pi}\cos \theta,
\end{equation}
\begin{equation}\label{A3}
c_{3L}=2\sqrt{2}\left[Z\phi_{3,1}^{(1)}+Z^2\phi_{3,1}^{(2)}\right],
\end{equation}
\begin{equation}\label{A4}
c_{2}=2\left\{\phi_{2,0}^{(0)}+
Z\left[\phi_{2,0}^{(1)}-C_{21}\cos \theta+\frac{\ln 2}{2}\phi_{2,1}^{(1)}\right]+Z^2 \phi_{2,0}^{(2)}
+a_{21}\cos\theta
\right\},
\end{equation}
\begin{equation}\label{A5}
c_{3}=2\sqrt{2}\left\{\phi_{3,0}^{(0)}+Z\left[\phi_{3,0}^{(1)}+\frac{\ln 2}{2}\phi_{3,1}^{(1)}\right]+
Z^2\left[\phi_{3,0}^{(2)}+\frac{\ln 2}{2}\phi_{3,1}^{(2)}\right]+Z^3 \phi_{3,0}^{(3)}+a_{21}(w_0+Zw_1)
\right\},
\end{equation}
where we denoted (for simplicity)
\begin{equation}\label{A6}
\phi_{k,p}^{(j)}\equiv\phi_{k,p}^{(j)}\left(\frac{\pi}{2},\theta\right).
\end{equation}
Functions $w_0\equiv w_0(\alpha,\theta)$ and $w_1 \equiv w_1(\alpha,\theta)$ will be defined later.
The AFC-components required for calculation of $c_1,c_{2L}$ and $c_{3L}$ represent simple analytic functions which can be found in Ref. \cite{LEZ4}. Therefore, we have presented coefficients $c_1$ and $c_{2L}$ in the final analytic form (see (\ref{A1})-(\ref{A2})), whereas the AFC-components
\begin{equation}\label{A7}
\phi_{3,1}^{(1)}=-\frac{(\pi-2)}{36\pi}(1+5\cos \theta)\sqrt{1-\cos \theta},~~~~~~~
\phi_{3,1}^{(2)}=\frac{(\pi-2)}{3\pi\sqrt{2}}\cos \theta,
\end{equation}
associated with $c_{3L}$, are presented separately, because they are included into Eq.(\ref{A5}) for $c_3$, as well.

It is seen that definitions (\ref{A4}) and (\ref{A5}) of the coefficients $c_2$ and $c_3$, respectively, contain both known and unknown AFC-components and parameters. In particular, the parameter
\begin{equation}\label{A8}
C_{21}=\frac{62+17\pi-48 G}{72\pi},
\end{equation}
where $G$ is the Catalan's constant, has been calculated in Ref. \cite{LEZ6}.
It characterizes the admixture of the unnormalized HH, $Y_{21}(\alpha,\theta)=\sin \alpha \cos \theta$ in the AFC-component $\phi_{2,0}^{(1)}(\alpha,\theta)$ defined by Eq.(22) from Ref. \cite{LEZ4}, whereas $a_{21}$ characterizes contribution of the mentioned HH into the physical AFC, $\phi_{2,0}(\alpha,\theta)$ associated with actual WF.

The AFC-components required for calculation of $c_2$ are:
\begin{equation}\label{A9}
\phi_{2,0}^{(0)}=\frac{1-2E}{12},~~~~~\phi_{2,0}^{(2)}=\frac{5}{6},~~~~~\phi_{2,1}^{(1)}=-\frac{\pi-2}{3\pi}\cos \theta,
\end{equation}
\begin{equation}\label{A10}
\phi_{2,0}^{(1)}=\frac{1}{6}\left\{1+2\left(1-\frac{\theta}{\pi}\right)\sin \theta-
\ln\left[2\left(1+\sin \frac{\theta}{2}\right)^2\right]\cos \theta-8\sin \frac{\theta}{2}
\right\}.
\end{equation}
 One should emphasize that expression in the RHS of Eq.(\ref{A10}) was obtained by simplification of  $\phi_{2,0}^{(1)}(\alpha,\theta)$ presented in Ref. \cite{LEZ4} (see Eq.(22) \cite{LEZ4}) and taken for $\alpha=\pi/2$.

Derivation of Eq.(\ref{A5}) for $c_3$ is based on the correct representation of the AFC, $\phi_{3,0}(\alpha,\theta)$ satisfying the FRR
\begin{equation}\label{B1}
(\Lambda^2-21)\phi_{3,0}=h_{3,0},
\end{equation}
where (see, e.g., Ref. \cite{LEZ4})
\begin{equation}\label{B2}
h_{3,0}=10\phi_{3,1}-2(V_0-Z V_1)\phi_{2,0}+2E\phi_{1,0}.
\end{equation}
Using definitions (\ref{109}) for $\phi_{3,1}$ and $\phi_{1,0}$, and also specific definition
\begin{equation}\label{B3}
\phi_{2,0}=\phi_{2,0}^{(0)}+Z\phi_{2,0}^{(1)}
+Z^2\phi_{2,0}^{(2)}+a_{21} \sin\alpha\cos\theta,
\end{equation}
been used for derivation of Eq.(\ref{A4}), we can present $h_{3,0}$ in the form:
\begin{equation}\label{B4}
h_{3,0}=\sum_{j=0}^3 Z^jh_{3,0}^{(j)}+a_{21}(\mathrm{h}_0+Z \mathrm{h}_1),
\end{equation}
where
\begin{equation}\label{B5}
h_{3,0}^{(0)}=-2V_0\phi_{2,0}^{(0)}+2E\phi_{1,0}^{(0)},
~~~~~~~~~~~h_{3,0}^{(3)}=2V_1\phi_{2,0}^{(2)},
\end{equation}
\begin{equation}\label{B6}
h_{3,0}^{(1)}=10 \phi_{3,1}^{(1)}-2V_0 \tilde{\phi}_{2,0}^{(1)}+2V_1\phi_{2,0}^{(0)}+2E\phi_{1,0}^{(1)},
\end{equation}
\begin{equation}\label{B7}
h_{3,0}^{(2)}=10 \phi_{3,1}^{(2)}-2V_0 \phi_{2,0}^{(2)}+2V_1\tilde{\phi}_{2,0}^{(1)},
\end{equation}
\begin{equation}\label{B8}
\mathrm{h}_0=-2V_0\sin\alpha\cos\theta,
\end{equation}
\begin{equation}\label{B9}
\mathrm{h}_1=2V_1\sin\alpha\cos\theta.
\end{equation}
Here, $\tilde{\phi}_{2,0}^{(1)}\equiv\phi_{2,0}^{(1)}-C_{21}\sin\alpha\cos\theta$ is the so called "pure" AFC (see Sec. \ref{S3}), whereas
the angular potentials $V_0$ and $V_1$ are defined by Eq.(\ref{112}).
Note that we omitted variables $(\alpha,\theta)$ of all functions in Eqs.(\ref{B1})-(\ref{B9}), for simplicity.

The simplest AFC-components required for calculation of $c_3$ are \cite{LEZ4}:
\begin{equation}\label{A11}
\phi_{3,0}^{(0)}=\frac{1}{72}\left[1-5E+(2-E)\cos \theta\right]\sqrt{1-\cos \theta},~~~~~~~~~~~
\phi_{3,0}^{(3)}=-\frac{7}{18\sqrt{2}}.
\end{equation}
To calculate $c_3$ we need also the AFC-components $\phi_{3,0}^{(1)}$ and $\phi_{3,0}^{(2)}$
which are rather complicated and therefore it requires detailed consideration.

\subsection{Calculation of $ \phi_{3,0}^{(1)}$}\label{SA1}

The RHS (\ref{B6}) can be presented in the form
\begin{equation}\label{A14}
h_{3,0}^{(1)}=h_{3,0}^{(1a)}+h_{3,0}^{(1b)}+h_{3,0}^{(1c)}+h_{3,0}^{(1d)}+h_{3,0}^{(1e)},
\end{equation}
where
\begin{eqnarray}\label{A15}
h_{3,0}^{(1a)}=-\frac{5(\pi-2)}{3\pi}\sqrt{1-\sin \alpha \cos \theta},~~~~
h_{3,0}^{(1b)}=\frac{1}{3}\sqrt{1+\sin \alpha}\left[\frac{1-2E}{\sin \alpha}+2(1-3E)\right],
~\nonumber~~\\
h_{3,0}^{(1c)}=\frac{25(\pi-2)}{18\pi}(1-\sin \alpha \cos \theta)^{-3/2},~~~~~~~~~~~~~~~~~~~~~~~~~~
\end{eqnarray}
\begin{equation}\label{A16}
h_{3,0}^{(1e)}=-\frac{2}{3}\sqrt{1+\sin \alpha},
\end{equation}
\begin{equation}\label{A17}
h_{3,0}^{(1d)}=-\frac{2\tilde{\phi}_{2,0}^{(1)}(\alpha,\theta)}{\sqrt{1-\sin \alpha \cos \theta}}.
\end{equation}
We omitted again variables $(a,\theta)$ of the $h$-functions for simplicity.
The AFC-components corresponding to the RHSs presented in Eq.(\ref{A15}) can be found in Ref. \cite{LEZ4}. For considered case $\alpha=\pi/2$, we easily obtain
\begin{eqnarray}\label{A18}
\phi_{3,0}^{(1a)}=\frac{5(\pi-2)\sqrt{2}}{36 \pi}\sin^3\left(\frac{\theta}{2}\right),~~~
\phi_{3,0}^{(1b)}=\frac{5E-2}{18\sqrt{2}},
~~~~~~~~~~~~\nonumber~~~~~~~~~~~~~\\
\phi_{3,0}^{(1c)}=\frac{5(\pi-2)}{864\pi\sqrt{1+\cos \theta}}\left\{
3(\pi-\theta)[1+2 \cos \theta+2\cos(2\theta)]+5\sin \theta(5\cos \theta-2)
\right\}.~~~~~~~~~~~~
\end{eqnarray}
It may seem that $\phi_{3,0}^{(1c)}$ is divergent at $\theta=\pi$. Actually, taking the relevant limit we obtain
$\phi_{3,0}^{(1c)}(\pi/2,\pi)=-5\sqrt{2}(\pi-2)/(27\pi)$.

The physical solution of the IFRR (\ref{107}) with the RHS $h_{3,0}^{(1e)}$ defined by Eq.(\ref{A16}) has not been determined previously. However, using the methods described in Ref. \cite{LEZ4}, we easily obtain:
\begin{equation}\label{A19}
\phi_{3,0}^{(1e)}(\alpha,\theta)=\frac{1}{18}\left[\sin^3(\alpha/2)+\cos^3(\alpha/2)\right].
\end{equation}
Whence, for $\alpha=\pi/2$ we obtain the result $\phi_{3,0}^{(1e)}=1/(18\sqrt{2})$ we are seeking for.

To calculate the AFC-component $\phi_{3,0}^{(1d)}$, first of all, let's present the RHS (\ref{A17}) of the corresponding IFRR in the form (like it was done for calculation of $c_2$):
\begin{equation}\label{A20}
h_{3,0}^{(1d)}=\mathrm{h}_{30}-C_{21}\mathrm{h}_{0},
\end{equation}
where function $\mathrm{h}_0\equiv \mathrm{h}_0(\alpha,\theta)$ is defined by Eqs.(\ref{B8}) and (\ref{112}), whereas
\begin{equation}\label{A22}
\mathrm{h}_{30}=-\frac{2\phi_{2,0}^{(1)}(\alpha,\theta)}{\sqrt{1-\sin \alpha\cos \theta}}.
\end{equation}
Using the methods described in Ref.\cite{LEZ4} (see also \cite{LEZ5}), we find the following physical solution of the IFRR (\ref{107}) with the RHS $\mathrm{h}_0$:
\begin{equation}\label{A23}
w_0(\alpha,\theta)=\frac{1}{12}(1+5\sin \alpha \cos\theta)\sqrt{1-\sin \alpha \cos\theta}.
\end{equation}
Whence, for $\alpha=\pi/2$ we obtain function $w_0=(1+5\cos\theta)\sqrt{1-\cos\theta}/12$ included into Eq.(\ref{A5}).

Function $\phi_{2,0}^{(1)}(\alpha,\theta)$ represents the AFC-component defined by Eq.(22) from Ref.\cite{LEZ4}. It is rather complicated function which particular case $\phi_{2,0}^{(1)}(\pi/2,\theta)$ is presented by Eq.(\ref{A10}).
The RHS (\ref{A22}) is too complicated to derive the physical solution $\chi_{30}$ of the corresponding IFRR in the form representing analytic (explicit) function of the hyperspherical angles $\alpha$ and $\theta$.
However, we can calculate numerically the value of  $\chi_{30}$ at any given point $\left\{\alpha_0,\theta_0\right\}$ of the hyperspherical angular space using the Green's function approach described in Sec. \ref{S3}. In particular, for the considered case with $k=3$ and $\alpha_0=\pi/2$, we obtain:
\begin{equation}\label{A24}
\chi_{30}\left(\frac{\pi}{2},\theta_0\right)=
\frac{1}{8\pi}\int_0^\pi d\alpha \sin^2\alpha\int_0^\pi d\theta \sin \theta~ \mathrm{h}_{30}(\alpha,\theta)
\int_0^\pi \frac{\cos(5\omega/2)}{\sin \omega}d\varphi,
\end{equation}
where
\begin{equation}\label{A25}
\cos \omega =\sin \alpha
\left(\cos \theta \cos \theta_0+\sin \theta \sin \theta_0 \cos \varphi\right).
\end{equation}
It is seen that for particular cases $\theta_0=0$ and $\theta_0=\pi$ one obtains $\cos \omega=\pm\sin\alpha \cos \theta$ which is independent on $\varphi$, whence the 3-dimensional integral (\ref{A24}) reduces to the 2-dimensional one.
Note that it is possible to reduce the 3-dimensional integral (\ref{A24}) to the 2-dimensional one in general case.
To this end, let's write down the trivial relation
\begin{equation}\label{A26}
g(\varphi)=\frac{\cos(5\omega/2)}{\sin \omega}=\frac{4\cos^2\omega-2\cos\omega-1}{\sqrt{2(1-\cos\omega)}},
\end{equation}
where (according to representation (\ref{A25}))
\begin{equation}\label{A27}
\cos\omega=x \cos\varphi+y.
\end{equation}
The required integral can be taken in the explicit form
\begin{equation}\label{A28}
\int_0^{\pi}g(\varphi)d\varphi=\frac{\sqrt{2}\left\{
[4x^2+4y(2-y)-1]\mathbb{K}\left(\frac{2x}{x-y+1}\right)-2(8y+1)(x-y+1)\mathbb{E}\left(\frac{2x}{x-y+1}\right)
\right\}}{3\sqrt{x-y+1}},
\end{equation}
where $\mathbb{K}(z)$ and $\mathbb{E}(z)$ are the complete elliptic integrals of the first and second kind, respectively.
Setting $x=\sin \alpha \sin \theta \sin\theta_0$ and $y=\sin \alpha \cos \theta \cos\theta_0$ in Eq.(\ref{A28}) we obtain the desired result.
Using Eqs.(\ref{A24})-(\ref{A28}) we have calculated $\chi_{30}(\pi/2,\theta_0)$ for
 $0\leq\theta\leq\pi$ with the step $h_{\theta_0}=\pi/8$.
The results are presented in Table \ref{T3} within an accuracy of 12 significant digits.
Note that the use of the explicit form (\ref{A28}) of the integral over $\varphi$ decreases the calculation time by the GF-approach \emph{substantially}.

The desired AFC-subcomponent can be calculated by the formula
\begin{equation}\label{A29}
\phi_{3,0}^{(1d)}=\chi_{30}-C_{21}w_{0}.
\end{equation}
And in general we obtain
\begin{equation}\label{A30}
\phi_{3,0}^{(1)}=\phi_{3,0}^{(1a)}+\phi_{3,0}^{(1b)}+\phi_{3,0}^{(1c)}+\phi_{3,0}^{(1d)}+\phi_{3,0}^{(1e)}.
\end{equation}

\subsection{Calculation of $ \phi_{3,0}^{(2)}$}\label{SA2}

Similar to considering $\phi_{3,0}^{(1)}$,
the RHS (\ref{B7}) of the relevant IFRR can be split into parts as follows:
\begin{equation}\label{A38}
h_{3,0}^{(2)}=h_{3,0}^{(2a)}+h_{3,0}^{(2b)}-\frac{4\pi}{5(\pi-2)}h_{3,0}^{(1a)}+h_{3,0}^{(2d)}+h_{3,0}^{(2e)},
\end{equation}
where
\begin{equation}\label{A39}
h_{3,0}^{(2a)}=-\frac{2}{3}\left(2\sqrt{1-\sin\alpha \cos \theta}+\frac{1}{\sqrt{1-\sin\alpha \cos \theta}}\right),~~
h_{3,0}^{(2b)}=\frac{5(\pi-2)}{3\pi}\sin\alpha\cos\theta\sqrt{1+\sin\alpha}~,
\end{equation}
\begin{equation}\label{A40}
h_{3,0}^{(2d)}=\frac{4\sqrt{1+\sin\alpha}}{\sin\alpha}~\tilde{\phi}_{2,0}^{(1)}(\alpha,\theta),
\end{equation}
\begin{equation}\label{A41}
h_{3,0}^{(2e)}=-\frac{\sin\alpha}{\sqrt{1-\sin\alpha\cos\theta}}.
\end{equation}
Accordingly, the AFC-component representing the solution of the IFRR (\ref{107}) with the RHS (\ref{A38}) can be written in the form
\begin{equation}\label{A42}\phi_{3,0}^{(2)}=\phi_{3,0}^{(2a)}+
\phi_{3,0}^{(2b)}-\frac{4\pi}{5(\pi-2)}\phi_{3,0}^{(1a)}+\phi_{3,0}^{(2d)}+\phi_{3,0}^{(2e)}.
\end{equation}
The AFC-components corresponding to the RHSs presented in Eq.(\ref{A39}) can be found in Ref. \cite{LEZ4}. For considered case $\alpha=\pi/2$, we obtain
\begin{equation}\label{A43}
\phi_{3,0}^{(2a)}=\frac{1}{18}(2+\cos\theta)\sqrt{1-\cos\theta},~~~~
\phi_{3,0}^{(2b)}=-\frac{(\pi-2)(5\pi+36)}{144\pi\sqrt{2}}\cos \theta.
\end{equation}
Function $\phi_{3,0}^{(1a)}\equiv\phi_{3,0}^{(1a)}(\pi/2,\theta)$ is defined by Eq.(\ref{A18}).

The solution $\phi_{3,0}^{(2e)}(\alpha,\theta)$ of the IFRR (\ref{107}) with the RHS (\ref{A41}) has been reported in Ref. \cite{LEZ5}. For the considered case $\alpha=\pi/2$ the desired result reduces to one-dimensional series of the form
\begin{equation}\label{A44}
\phi_{3,0}^{(2e)}\left(\frac{\pi}{2},\theta\right)=\sum_{l=0}^\infty
\left[\frac{2l(l+1)\left(H_{\frac{l-1}{2}}-H_{\frac{l}{2}}\right)-2l-1}
{2\sqrt{2}(2l-3)(2l+1)(2l+5)}\right]P_l(\cos \theta),
\end{equation}
where the Harmonic number $H_z$ is related to the Euler constant $\gamma$ and the digamma function $\psi_0(z+1)$ by $H_z=\gamma+\psi_0(z+1)$, and where $P_l(z)$ are the Legendre polynomials.
We cannot provide summation of the infinite series (\ref{A44}) in general analytic form. However, we can calculate the sum of the infinite series (\ref{A44}) for any given angle $\theta$. The equivalent numerical results can be obtained by the GF approach (see Sec.\ref{S3}) using the RHS (\ref{A41}).
Moreover, for the cases $\theta=0,\pi,\pi/2$ under consideration, it is possible to provide summation of the infinite series (\ref{A44}) in the explicit (analytic) form. In particular, we obtain:
\begin{equation}\label{A45}
\phi_{3,0}^{(2e)}\left(\frac{\pi}{2},0\right)=\frac{5-6 G}{12\sqrt{2}}.~~~~~~~~~~~
\phi_{3,0}^{(2e)}\left(\frac{\pi}{2},\pi\right)=-\frac{\pi^2-4}{32\sqrt{2}}
\end{equation}
Also, the nontrivial operations by Wolfram  \emph{Mathematica} yield the following result:
\begin{eqnarray}\label{A46}
\phi_{3,0}^{(2e)}\left(\frac{\pi}{2},\frac{\pi}{2}\right)=\frac{1}{45\sqrt{2}}
\left\{\frac{15(6\ln 2-1)}{8\sqrt{\pi}}\Gamma\left(\frac{5}{4}\right)^2+
~_2F_1\left(\frac{1}{4},\frac{3}{2};\frac{13}{4};-1\right)+
\right.
~~~~~~~~~\nonumber~~\\
\left.
3\left[_3F_2^{(\left\{0,0,0\right\},\left\{0,1\right\},0)}\left(\frac{1}{4},\frac{5}{2},\frac{3}{2};\frac{13}{4},\frac{3}{2};-1\right)
-~_3F_2^{(\left\{0,0,0\right\},\left\{0,1\right\},0)}\left(\frac{1}{4},\frac{5}{2},2;\frac{13}{4},2;-1\right)
\right]
\right\}
~~~~~~~~~\nonumber~~\\
\simeq 0.0583~734~256~330~535~678~.~~~~~~~~~~~~~~~~~~~
\end{eqnarray}
Here, $_2F_1(...)$ is the Gauss hypergeometric function, and $_3F_2^{(\left\{0,0,0\right\},\left\{0,1\right\},0)}(a_1,a_2,a_3;b_1,b_2;z)$ is the derivative of the corresponding hypergeometric function in respect to the parameter $b_2$.
In spite of representation (\ref{A46}) is rather complicated, it enables us to calculate $\phi_{3,0}^{(2e)}(\pi/2,\pi/2)$ with any predetermined accuracy (e.g, 18 significant digits are presented).

It should be marked that the results (\ref{A45}) can be obtained by the GF-formula (\ref{104}), as well. For the cases $\theta'=0,\pi$ representation (\ref{106}) for $\cos\omega$ is independent on auxiliary angle $\varphi$, and the relevant two-dimension integrals can be taken in the explicit form.

The only undetermined (currently) AFC-component is $\phi_{3,0}^{(2d)}$. The  RHS $h_{3,0}^{(2d)}$ of the corresponding IFRR is defined by Eq.(\ref{A40}) which includes the "pure" AFC-component $\tilde{\phi}_{2,0}^{(1)}(\alpha,\theta)$.
Setting, as previously, $\tilde{\phi}_{2,0}^{(1)}=\phi_{2,0}^{(1)}-C_{21}\sin\alpha\cos\theta$, we obtain
\begin{equation}\label{A47}
 h_{3,0}^{(2d)}=\mathrm{h}_{31}-C_{21}\mathrm{h}_1,
\end{equation}
and accordingly
\begin{equation}\label{A48}
 \phi_{3,0}^{(2d)}=\chi_{31}-C_{21}w_1,
\end{equation}
The RHS $\mathrm{h}_1=4\cos\theta\sqrt{1+\sin\alpha}$ is defined by Eqs.(\ref{B9}) and (\ref{112}), whereas
\begin{equation}\label{A49}
 \mathrm{h}_{31}=\frac{4\sqrt{1+\sin\alpha}}{\sin\alpha}~\phi_{2,0}^{(1)}(\alpha,\theta).
\end{equation}
Using the methods described in Ref.\cite{LEZ4}, we derive the physical solution of the IFRR (\ref{107}) with the RHS, $\mathrm{h}_1$ in the form
\begin{equation}\label{A50}
 w_1(\alpha,\theta)=-\frac{1}{2}\sin\alpha\cos\theta\sqrt{1+\sin\alpha}.
\end{equation}
Whence, for $\alpha=\pi/2$ we obtain function $w_1=-\cos\theta/\sqrt{2}$ included into Eqs.(\ref{A5}) and (\ref{A48}).

The RHS (\ref{A49}) is too complicated due to the presence of the AFC-component $\phi_{2,0}^{(1)}(\alpha,\theta)$ defined by Eq.(22) from Ref.\cite{LEZ4}.
 We cannot derive the physical solution $\chi_{31}(\alpha,\theta)$ of the corresponding IFRR in the analytic form.
However, we can calculate numerically the value of  $\chi_{31}$ at any given point $\left\{\alpha_0,\theta_0\right\}$ using the Green's function approach.
In particular, for the considered case with $k=3$ and $\alpha_0=\pi/2$, we obtain:
\begin{equation}\label{A51}
\chi_{31}\left(\frac{\pi}{2},\theta_0\right)=
\frac{1}{8\pi}\int_0^\pi d\alpha \sin^2\alpha\int_0^\pi d\theta \sin \theta~ \mathrm{h}_{31}(\alpha,\theta)
\int_0^\pi \frac{\cos(5\omega/2)}{\sin \omega}d\varphi,
\end{equation}
where the angle $\omega$ is defined by Eq.(\ref{A25}). To simplify calculations one can apply the representations (\ref{A26})-(\ref{A28}) that were described in the previous Subsection.
Using Eq. (\ref{A51}) we have calculated $\chi_{31}(\pi/2,\theta_0)$ for $0\leq\theta_0\leq\pi$ with step $h_{\theta_0}=\pi/8$.
The results are presented in Table \ref{T3} within accuracy of 12 significant digits.

\newpage

\begin{table}
\caption{The expectation values $S(\theta,Z)$ defined by Eqs.(\ref{17})-(\ref{17a}) and calculated by the PLM with number of shells $\Omega$ indicated in parentheses. }
\begin{tabular}{|c|c|c|c|c|c|}
\hline
$\theta \setminus Z(\Omega)$ & $1(30)$ & $2(25)$ & $3(30)$& $4(26)$& $5(27)$ \tabularnewline
\hline
$0$      & 0.002 738 0 & 0.106 345 4 & 0.533 722 5 & 1.522 895 7 & 3.312 442 5 \tabularnewline
\hline
$\pi/8$  & 0.003 747 3 & 0.122 640 2 & 0.585 410 0 & 1.630 718 3 & 3.497 271 1 \tabularnewline
\hline
$\pi/4$  & 0.004 838 1 & 0.137 582 6 & 0.630 549 0 & 1.722 694 2 & 3.652 771 7 \tabularnewline
\hline
$3\pi/8$ & 0.005 937 0 & 0.150 803 8 & 0.668 982 8 & 1.799 576 6 & 3.781 354 2 \tabularnewline
\hline
$\pi/2$  & 0.006 968 1 & 0.162 010 4 & 0.700 595 5 & 1.861 910 8 & 3.884 729 4 \tabularnewline
\hline
$5\pi/8$ & 0.007 859 1 & 0.170 967 9 & 0.725 290 6 & 1.910 075 4 & 3.964 094 2 \tabularnewline
\hline
$3\pi/4$ & 0.008 562 0 & 0.177 495 5 & 0.742 987 2 & 1.944 323 2 & 4.020 265 7 \tabularnewline
\hline
$7\pi/8$ & 0.009 121 9 & 0.181 463 9 & 0.753 625 4 & 1.964 813 8 & 4.053 771 8 \tabularnewline
\hline
$\pi$    & 0.009 413 6 & 0.182 795 3 & 0.757 174 1 & 1.971 634 7 & 4.064 908 8 \tabularnewline
\hline
\end{tabular}
\label{T1}
\end{table}

\begin{table}
\caption{The characteristic angles $\theta_v,\theta_t,\theta_{cr}$ and $\theta_{vir}$ (in the units of $\pi$) of the
expectation values $\langle T^{(\theta)}\rangle$ and $\langle V^{(\theta)}\rangle$
for the ground state of the two-electron atomic systems (with nucleus charge $Z$) in the "electrons-on-sphere" configuration.
}
\begin{tabular}{|c|c|c|c|c|}
\hline
$Z\setminus \theta
$ & $\theta_{v}$ & $\theta_{cr}$ & $\theta_{t}$ & $\theta_{vir}$\tabularnewline
\hline
\hline
1 &0.16086  & 0.17656 & 0.19622  &0.25707 \tabularnewline
\hline
2 &0.079786  & 0.088706   & 0.099978 &0.13501   \tabularnewline
\hline
3 &0.053113  & 0.059492 &0.067652  &0.093610  \tabularnewline
\hline
4 &0.039815 &0.044790 & 0.051205 & 0.071956\tabularnewline
\hline
5 & 0.031844  & 0.035924& 0.041214 & 0.058528 \tabularnewline
\hline
\end{tabular}
\label{T2}
\end{table}

\begin{table}
\caption{The AFC-subcomponents $\chi_{30}(\pi/2,\theta)$, $\chi_{31}(\pi/2,\theta)$ and $\phi_{3,0}^{(2e)}(\pi/2,\theta)$ calculated by the Green's function method.}
\begin{tabular}{|c|c|c|c|}
\hline
$\theta \setminus \chi$ & $\chi_{30}$ & $\chi_{31}$&$\phi_{3,0}^{(2e)}$ \tabularnewline
\hline
$0$     & -0.0618 939 408 156  & 0.0759 589 476 113&-0.0292 149 159 937 435 958  \tabularnewline
\hline
$\pi/8$ & -0.0543 599 956 237 &  0.0756 722 185 180& 0.0359 428 375 585 742   \tabularnewline
\hline
$\pi/4$ & -0.0477 844 616 884  & 0.0889 019 470 391 &0.0793 908 418 427 414 \tabularnewline
\hline
$3\pi/8$ &-0.0441 969 089 641   & 0.131 424 424 880  &  0.0875 307 027 058 007    \tabularnewline
\hline
$\pi/2$ & -0.0443 241 463 159  &  0.206 732 030 685 &0.0583 734 256 330 535 678 \tabularnewline
\hline
$5\pi/8$ & -0.0475 018 819 969 &0.303 933 756 159 &0.00229 147 250 603 431 \tabularnewline
\hline
$3\pi/4$&-0.0520 306 782 191   & 0.401 135 276 615  & -0.0614 271 771 906 796 \tabularnewline
\hline
$7\pi/8$& -0.0558 555 609 027 & 0.472 955 896 676 & -0.111 048 941 538 468 \tabularnewline
\hline
$\pi$   & -0.0573 377 187 869  &  0.499 412 395 985 & -0.129 701 158 590 396 545 \tabularnewline
\hline
\end{tabular}
\label{T3}
\end{table}

\begin{table}
\caption{Parameters of the model WF (\ref{122}) for the ground state of the helium atom. Angle $\theta$ characterises the mutual arrangement of the electrons, whereas $\mathcal{R}$ characterizes the respective accuracy of the model WF under consideration.}
\begin{tabular}{|c|c|c|c|c|}
\hline
$\theta  $ & $b$ & $\kappa$&$\sigma$& $\mathcal{R}\times10^3$\tabularnewline
\hline
$0$      & 3.341 674 693 & 3.213 986 461 & 2.031 971 222 & 1.11 \tabularnewline
\hline
$\pi/8$  & 3.263 200 206 & 2.690 919 377 & 0.784 041 511 & 2.62 \tabularnewline
\hline
$\pi/4$  & 3.318 582 761 & 2.178 279 213 & 1.675 651 366 & 1.33 \tabularnewline
\hline
$3\pi/8$ & 3.341627+0.013514 $i$ & 2.026019-0.014753 $i$ & 3.212711+1.302211 $i$ & 0.35\tabularnewline
\hline
$~\pi/2$ & 2.995207-0.105835 $i$ &         1             &-0.290804-0.116110 $i$ & 6.55\tabularnewline
\hline
$5\pi/8$ & 3.374 788 170 & 1.875 718 550 & 1.659 602 319 & 1.01 \tabularnewline
\hline
$3\pi/4$ & 3.365 781 506 & 1.879 564 152 & 2.405 106 480 & 0.64 \tabularnewline
\hline
$7\pi/8$ & 3.357 052 601 & 2.880 748 298 & 1.882 725 651 & 0.08 \tabularnewline
\hline
$\pi$    & 3.352 074 330 & 1.885 859 328 & 3.084 565 080 & 0.33 \tabularnewline
\hline
\end{tabular}
\label{T4}
\end{table}

\begin{table}
\caption{Parameters of the model WF (\ref{131}), which being dependent on the nucleus charge $Z$ and the energy $E$ only,
were calculated by Eqs.(\ref{137})-(\ref{143}) for the ground state of the helium atom.
The parameters $\theta$ and $\mathcal{R}$ are the same as in Table \ref{T4}.}
\begin{tabular}{|c|c|c|c|c|c|}
\hline
$\theta  $ & $b$ & $\kappa$&$\sigma$& $\lambda$ &$\mathcal{R}\times10^3$\tabularnewline
\hline
$0$      & 3.940089-0.296849 $i$& 2.103769+0.0288802 $i$& 2.978964-0.662703 $i$ & 0.866252+0.122242 $i$ &1.19 \tabularnewline
\hline
$\pi/8$  & 3.729925-0.282081 $i$& 2.099957+0.0235201 $i$& 2.932936-0.587921 $i$ & 0.876545+0.126375 $i$ &1.79 \tabularnewline
\hline
$\pi/4$  & 3.554052-0.140415 $i$& 2.078883+0.0082375 $i$& 2.954381-0.278769 $i$ & 0.906646+0.069469 $i$ &0.67 \tabularnewline
\hline
$3\pi/8$ & 3.213 689 360        & 2.045 887 813         & 2.694 550 196         & 1.064 178 962         &1.88 \tabularnewline
\hline
$\pi/2$  & 3.485 261 332        & 2                     & 3.898 339 186         & 0.889 610 508         &1.16 \tabularnewline
\hline
$5\pi/8$ & 3.461 788 376        & 1.980 769 532         & 4.726 941 733         & 0.879 938 613         &2.18 \tabularnewline
\hline
$3\pi/4$ & 3.432 157 870        & 1.973 940 307         & 5.565 367 736         & 0.871 616 954         &1.70 \tabularnewline
\hline
$7\pi/8$ & 3.426 113 170        & 1.969 923 115         & 5.973 588 656         & 0.860 012 149         &1.65 \tabularnewline
\hline
$\pi$    & 3.352 267 303        & 1.892 703 054         & 3.191 143 089         & 0.993 443 303         &1.11 \tabularnewline
\hline
\end{tabular}
\label{T5}
\end{table}


\begin{thebibliography}{99}

\bibitem{KES} N. R. Kestner and O. Sinanoglu, Study of Electron Correlation in Helium-Like Systems Using an
Exactly Soluble Model, Phys. Rev. 128 (1962) 2687-2692.

\bibitem{KAI} S. Kais, D. R. Herschbach and R. D. Levine, Dimensional scaling as a symmetry operation,  J. Chem. Phys. 91 (1989) 7791-7796.

\bibitem{TAU} M. Taut, Two electrons in an external oscillator potential: Particular analytic solutions of a Coulomb correlation problem, Phys. Rev. A 48 (1993) 3561-3566.

\bibitem{ALA} A. Alavi, Two interacting electrons in a box: An exact diagonalization study, J. Chem. Phys. 113 (2000) 7735-7745.

\bibitem{TH1} D. C. Thompson and A. Alavi, Two interacting electrons in a spherical box: An exact diagonalization study, Phys. Rev. B 66 (2002) 235118-11.

\bibitem{TH2} D. C. Thompson and A. Alavi, A comparison of Hartree-Fock and exact diagonalization solutions for a model two-electron
    system, J. Chem. Phys. 122 (2005) 124107-7.

\bibitem{EZR1} G. S. Ezra and R. S. Berry, Correlation of two particles on a sphere, Phys. Rev. A 25 (1982) 1513-1527.

\bibitem{EZR2} G. S. Ezra and R. S. Berry, Quantum states of two particles on concentric spheres, Phys. Rev. A 28 (1983) 1989-2000.

\bibitem{OJH} P. C. Ojha and R. S. Berry, Angular correlation of two electrons on a sphere, Phys. Rev. A 36 (1987) 1575-1585.

\bibitem{HIN} R. J. Hinde and R. S. Berry, Correlation of two weakly attractive particles on a sphere, Phys. Rev. A 42 (1990) 2259-2266.

\bibitem{WAR} J. W. Warner and R. S. Berry, Hund's rule, Nature (London) 313 (1985) 160.


\bibitem{SEI1} M. Seidl, Adiabatic connection in density-functional theory: Two electrons on the surface of a sphere, Phys. Rev. A 75 (2007) 062506-11.

\bibitem{LO1} P.-F. Loos and P. M. Gill, Ground state of two electrons on a sphere, Phys. Rev. A 79 (2009) 062517-8.

\bibitem{LEZ1} E. Z. Liverts and N. Barnea, $S$-states of helium-like ions, Comp. Phys. Comm. 182 (2011) 1790-1795.

\bibitem{LEZ2} E. Z. Liverts and N. Barnea, Three-body systems with Coulomb interaction. Bound and quasi-bound $S$-states, Comp. Phys. Comm. 184 (2013) 2596-2603.

\bibitem{LEZ3} E. Z. Liverts, R. Krivec and N. Barnea, Collinear configuration of the helium atom and two-electron ions, Annals of Physics 422 (2020) 168306-17.

\bibitem{FR1} A. M. Frolov and V. H. Smith Jr., Exponential representation in the Coulomb three-body problem, J. Phys. B 37 (2004) 2917-2932.

\bibitem{FR2} A. M. Frolov, Field shifts and lowest order QED corrections for the ground  $1^1S$ and $2^3S$ states of the helium atoms, J. Chem. Phys. 126 (2007) 104302-10.

\bibitem{FR3} A. M. Frolov, On the Q-dependence of the lowest-order QED corrections and other properties of the ground $1^1S$-states in the two-electron ions, Chem. Phys. Lett. 638 (2015) 108-115.

\bibitem{GOT} J. E. Gottschalk and E. N. Maslen, Coordinate systems and analytic expansions for three-body atomic wavefunctions: III. Derivative continuity via solutions to Laplace's equation, J. Phys. A: Math. Gen. 20 (1987) 2781-2803.

\bibitem{FOCK} V. A. Fock, On the Schrodinger Equation of the Helium Atom,  \emph{Izv. Akad. Nauk SSSR, Ser. Fiz.} 18 (1954) 161-174;
   "V. A. Fock - Selected Works: Quantum Mechanics and Quantum Field Theory ", edited by L. D. Fadeev, L. A. Khalfin, I. V. Komarov, London, N-Y, Washington D.C., 2004, p.525.

\bibitem{MORG1} J. D. Morgan III, Convergence properties of Fock's expansion for $S$-state eigenfunctions of the helium atom,  Theor. Chim. Acta 69 (1986) 181-223.

\bibitem{LEZ4} E. Z. Liverts and N. Barnea,  Angular Fock coefficients. Refinement and further development,
Phys. Rev. A 92 (2015) 042512-21.

 \bibitem{AB1} P. C. Abbott and E. N. Maslen, Coordinate systems and analytic expansions for three-body atomic wavefunctions:
I. Partial summation for the Fock expansion in hyperspherical coordinates, J. Phys. A: Math. Gen. 20 (1987) 2043-2075.


\bibitem{LEZ5} E. Z. Liverts,  Analytic calculation of the edge components of the angular Fock coefficients,
Phys. Rev. A 94 (2016) 022504-13.

\bibitem{LEZ6} E. Z. Liverts,  Two-particle atomic coalescences: Boundary conditions for the Fock coefficient components,
Phys. Rev. A 94 (2016) 022506-13.

\bibitem{LEZ7} E. Z. Liverts and N. Barnea,  The Green's function approach to the Fock expansion calculations of two-electron atoms,
J. Phys. A: Math. Theor. 51 (2018) 085204 (13pp).

\bibitem{DRK} G. W. F. Drake, High Precision Calculations for Helium, Sec. 11 in "Atomic, Molecular, and Optical Physics Handbook", edited by G. W. F. Drake, AIP Press, New York, 1996.

\bibitem{MYERS} C. R. Myers, C. J. Umrigar, J. P. Sethna J P and J. D. Morgan III, Fock's expansion, Kato's cusp conditions, and the exponential ansatz, Phys. Rev. A 44 (1991) 5537-5546.


\end{thebibliography}
\end{document}